\documentclass{article}
\usepackage{spconf,amsmath,graphicx}
\usepackage[dvipsnames]{xcolor}
\usepackage{diagbox}
\usepackage{hyperref}
\usepackage{scrextend}
\title{\vspace{-10mm}Upsampling layers for music source separation\vspace{-0mm}}

\name{Jordi Pons, Joan Serrà, Santiago Pascual, Giulio Cengarle, Daniel Arteaga, Davide Scaini\vspace{-2mm}}

\address{Dolby Laboratories\vspace{-3mm}}

\begin{document}
\ninept
\maketitle
\begin{abstract} 

\noindent Upsampling artifacts are caused by problematic upsampling layers and due to spectral replicas that emerge while upsampling.
Also, depending on the used upsampling layer, such artifacts can either be tonal artifacts (additive high-frequency noise) or filtering artifacts (substractive, attenuating some bands). 
In this work we investigate the practical implications of having upsampling artifacts in the resulting audio, by studying how different artifacts interact and assessing their impact on the models' performance. To that end, we benchmark a large set of upsampling layers for music source separation: different transposed and subpixel convolution setups, different interpolation upsamplers (including two novel layers based on stretch and sinc interpolation), and different wavelet-based upsamplers (including a novel learnable wavelet layer). Our results show that filtering artifacts, associated with interpolation upsamplers, are perceptually preferrable, even if they tend to achieve worse objective scores.

\end{abstract}

\vspace{-2mm}
\section{Introduction}
\vspace{-1mm}

Upsampling layers are widely used for audio synthesis and can introduce undesired upsampling artifacts~\cite{donahue2018adversarial, stoller2018wave, pons2020upsampling, pandey2020densely, kumar2019melgan}, which can be categorized as filtering and tonal artifacts~\cite{pons2020upsampling}. Filtering artifacts attenuate some bands and are known to ``de-emphasize high-end frequencies"~\cite{pons2020upsampling}, while tonal artifacts introduce additive periodic noise percieved as a ``high-frequency buzzing noise"~\cite{stoller2018wave} that ``can be devastating to audio generation results"~\cite{donahue2018adversarial}.
Since sorting out such artifacts is instrumental for
the development of high-fidelity neural audio synthesizers, 
our work aims at investigating current upsampling layers to advance our understanding of those: 
transposed convolutions~\cite{donahue2018adversarial,kumar2019melgan,defossez2019music}, interpolation upsamplers~\cite{ stoller2018wave,gritsenko2020spectral}, subpixel convolutions~\cite{pandey2020densely}, and wavelet-based upsamplers~\cite{nakamura2020time}.
Transposed and subpixel convolutions can introduce tonal artifacts~\cite{donahue2018adversarial, stoller2018wave,kumar2019melgan}, whereas interpolation and wavelet-based upsamplers can produce filtering artifacts~\cite{pons2020upsampling}. 
In addition, recent research describes that tonal and filtering artifacts interact with the spectral replicas introduced via the bandwidth extension performed by each upsampling layer~\cite{pons2020upsampling}.
In light of that, which upsampling layers are preferrable? 
In section~2, we discuss their characteristics and how they interact with spectral replicas.
Given that the role of spectral replicas in neural networks was recently introduced, which is its practical impact? We discuss its implications, both theoretically and empirically, in sections 2.5, 3 and~4.
Further, in sections 3 and 4,  we extensively benchmark 
the above listed upsampling layers to understand their behavior, and 
investigate two additional strategies tailored to mitigate upsampling artifacts:~(i)~employing post-processing networks, as an ``a posteriori" mechanism to palliate the artifacts introduced by upsampling layers~\cite{donahue2018adversarial,dhariwal2020jukebox}; and~(ii)~using normalization layers, as a way to mitigate the spectral replicas of signal offsets---which we find to be a strong source of additional tonal artifacts~(see section 2.5). Hence, in short, our main contributions are to keep advancing our understanding of how to use upsampling layers for audio synthesis; and to experiment with upsampling layers that, to the best of our knowledge, have never been used for audio synthesis:  stretch and sinc interpolation layers, and \mbox{learnable wavelet layers (details in section 2).}

\begin{figure*}[h]
	\vfill \vspace{-11mm}
	\hspace{18mm}
	\begin{minipage}[b]{0.18\linewidth}
		\centering
		\centerline{\includegraphics[width=1.1\linewidth]{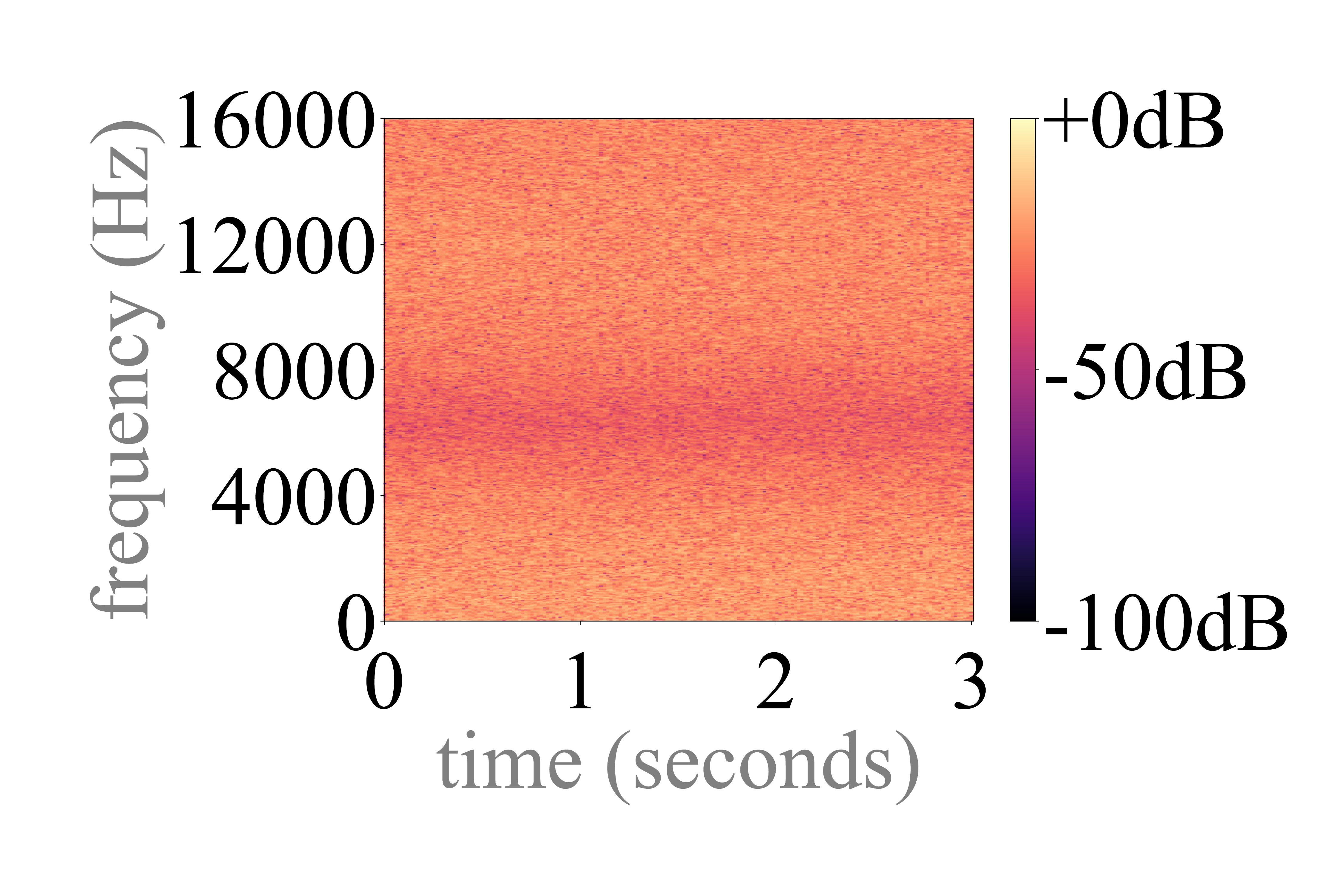}}
		\vspace{-3mm}
		{(a) Transposed CNN:\\no overlap (length=8)}\medskip
	\end{minipage}	
	\begin{minipage}[b]{0.18\linewidth}
		\centering
		\centerline{\includegraphics[width=1.1\linewidth]{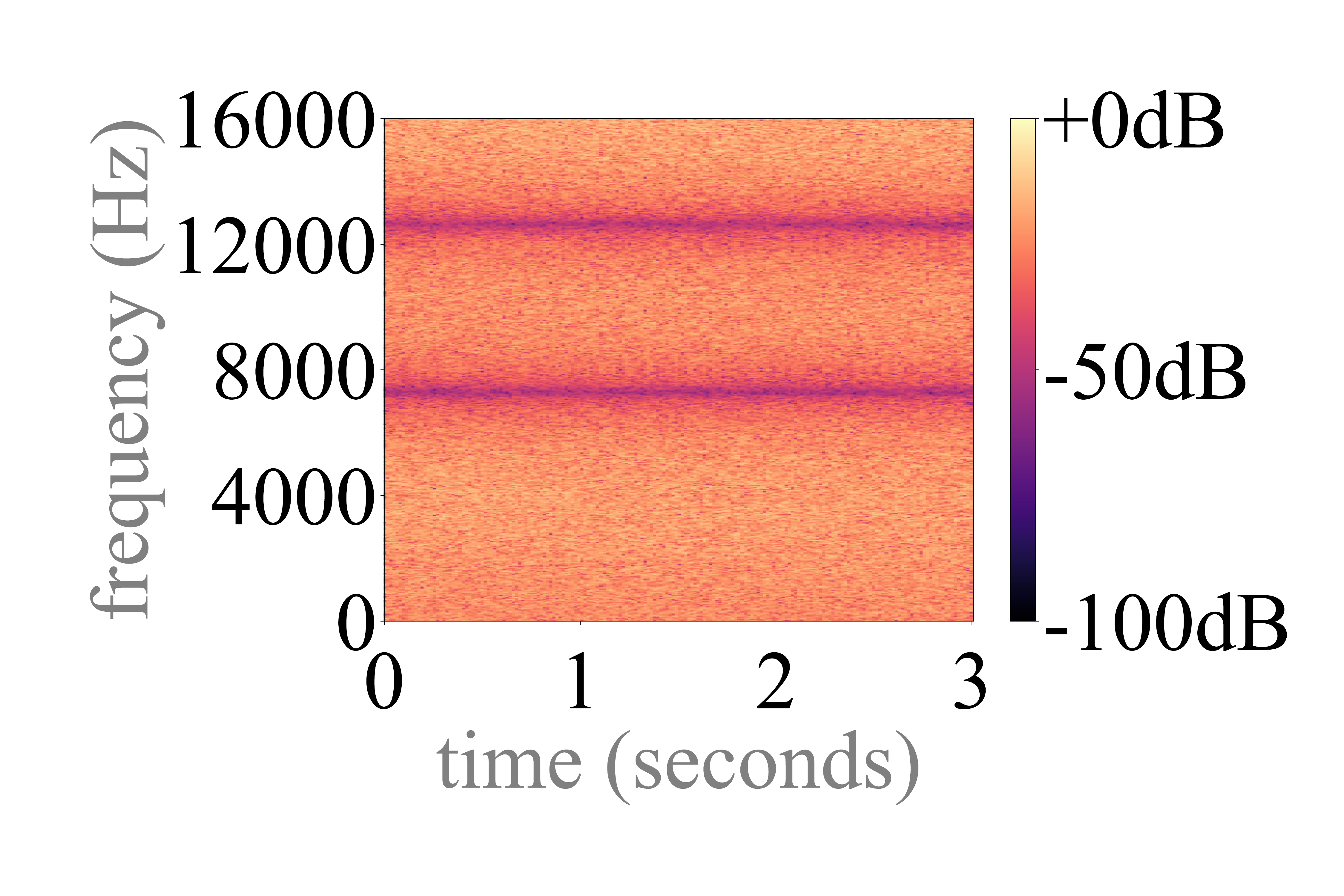}}
		\vspace{-3mm}
		{(b) Transposed CNN:\\\mbox{partial overlap (length=9)}}\medskip
	\end{minipage}
	\begin{minipage}[b]{0.18\linewidth}
		\centering
		\centerline{\includegraphics[width=1.1\linewidth]{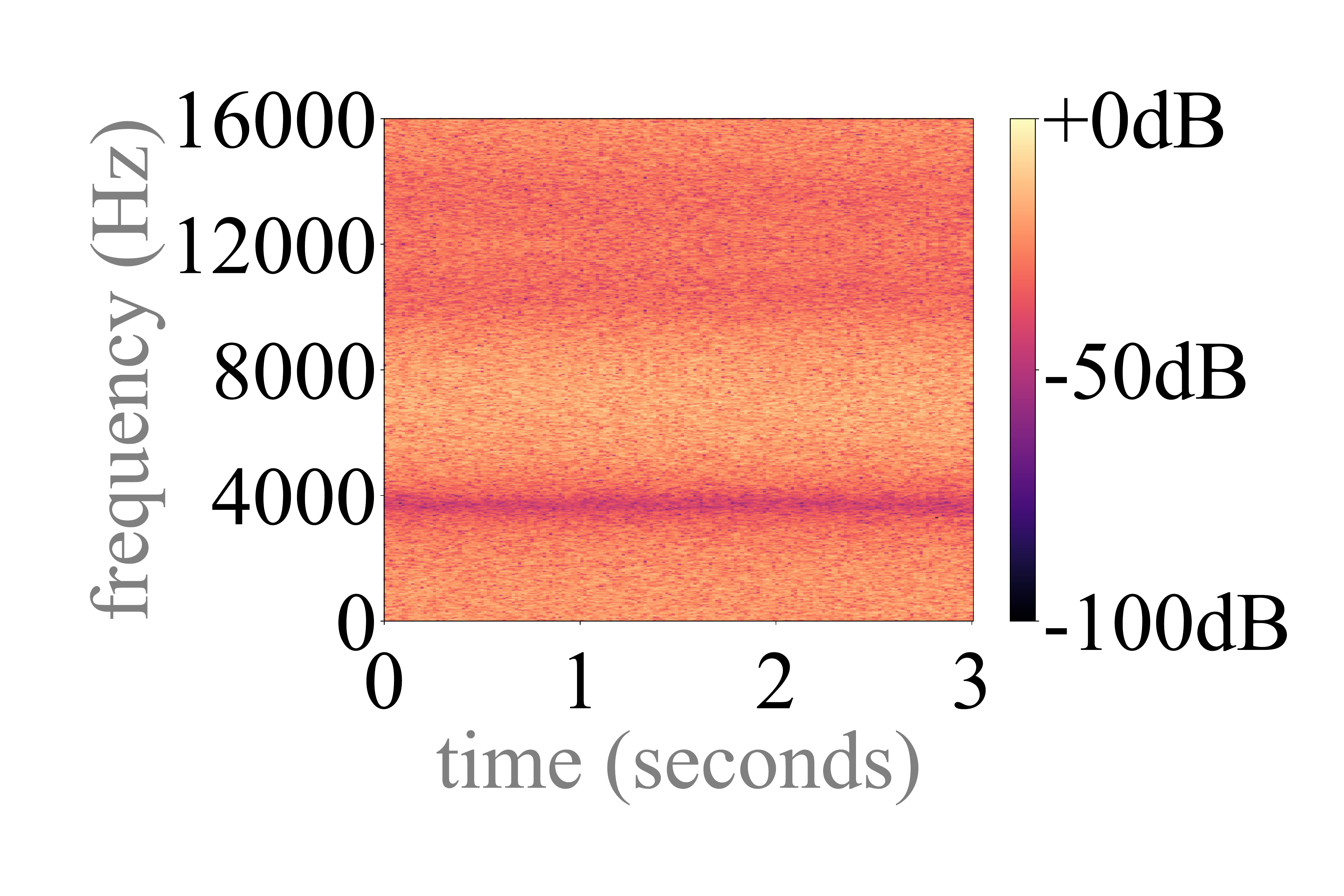}}
		\vspace{-3mm}
		{(c) Transposed CNN:\\full overlap (length=4)}\medskip
	\end{minipage}
	\begin{minipage}[b]{0.18\linewidth}
		\centering
		\centerline{\includegraphics[width=1.1\linewidth]{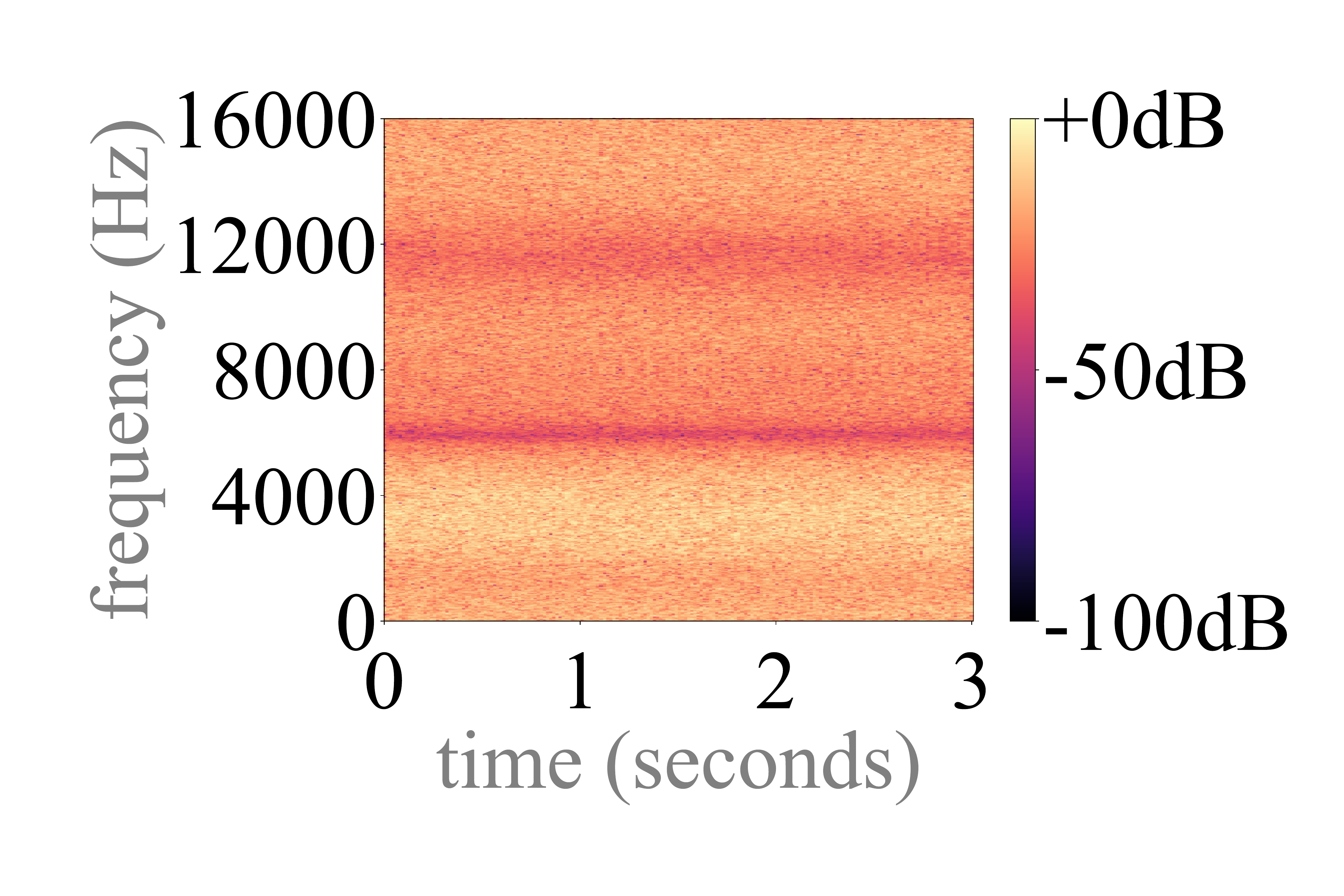}}
		\vspace{-3mm}
		{(d) Subpixel CNN\\(filter length=9)}\medskip
	\end{minipage}
	\vfill \vspace{-4mm}
		\hspace{18mm}
	\begin{minipage}[b]{0.18\linewidth}
		\centering
		\centerline{\includegraphics[width=1.1\linewidth]{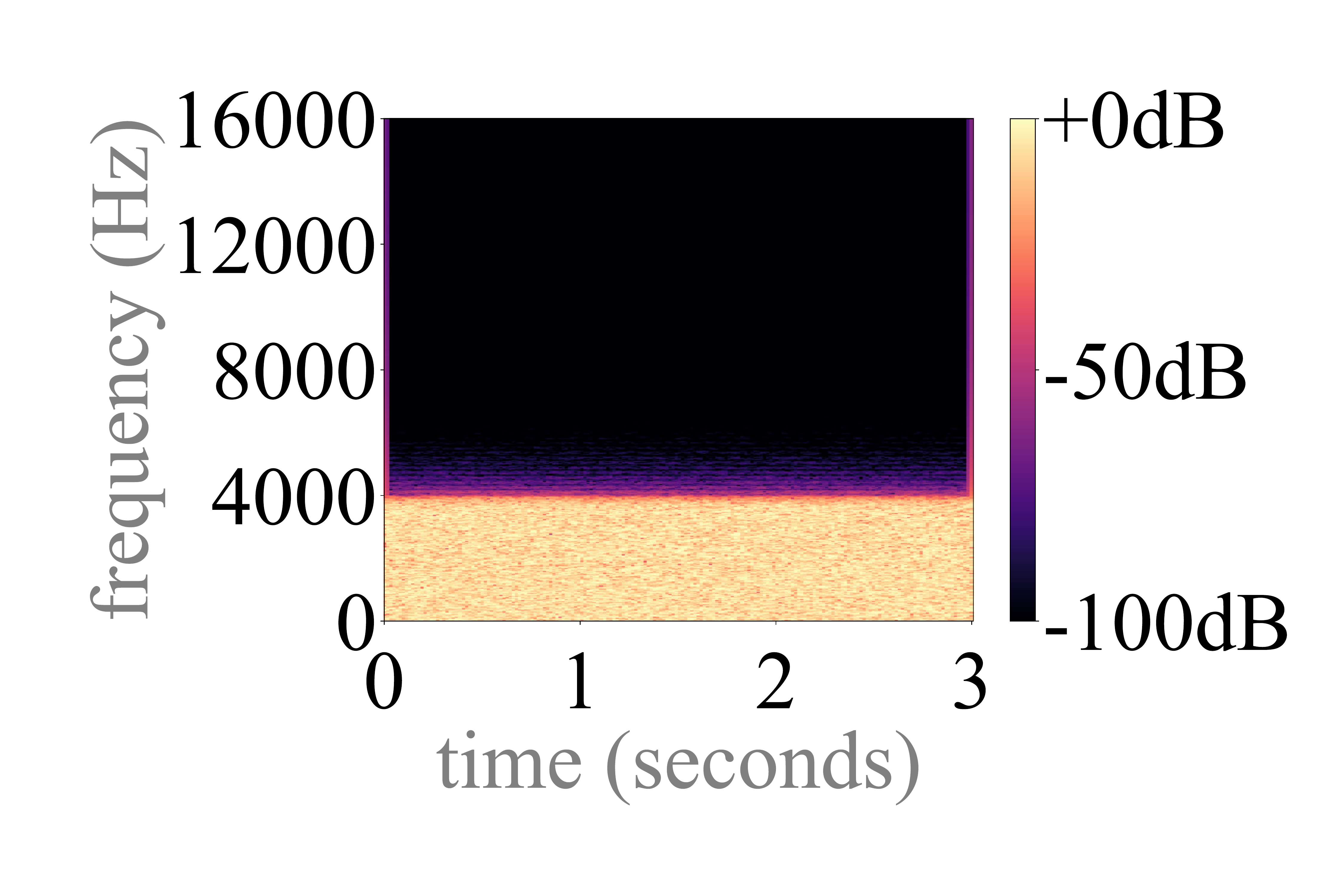}}
		\vspace{-3mm}
		{(e) Interpolation:\\sinc}\medskip
	\end{minipage}	
	\begin{minipage}[b]{0.18\linewidth}
		\centering
		\centerline{\includegraphics[width=1.1\linewidth]{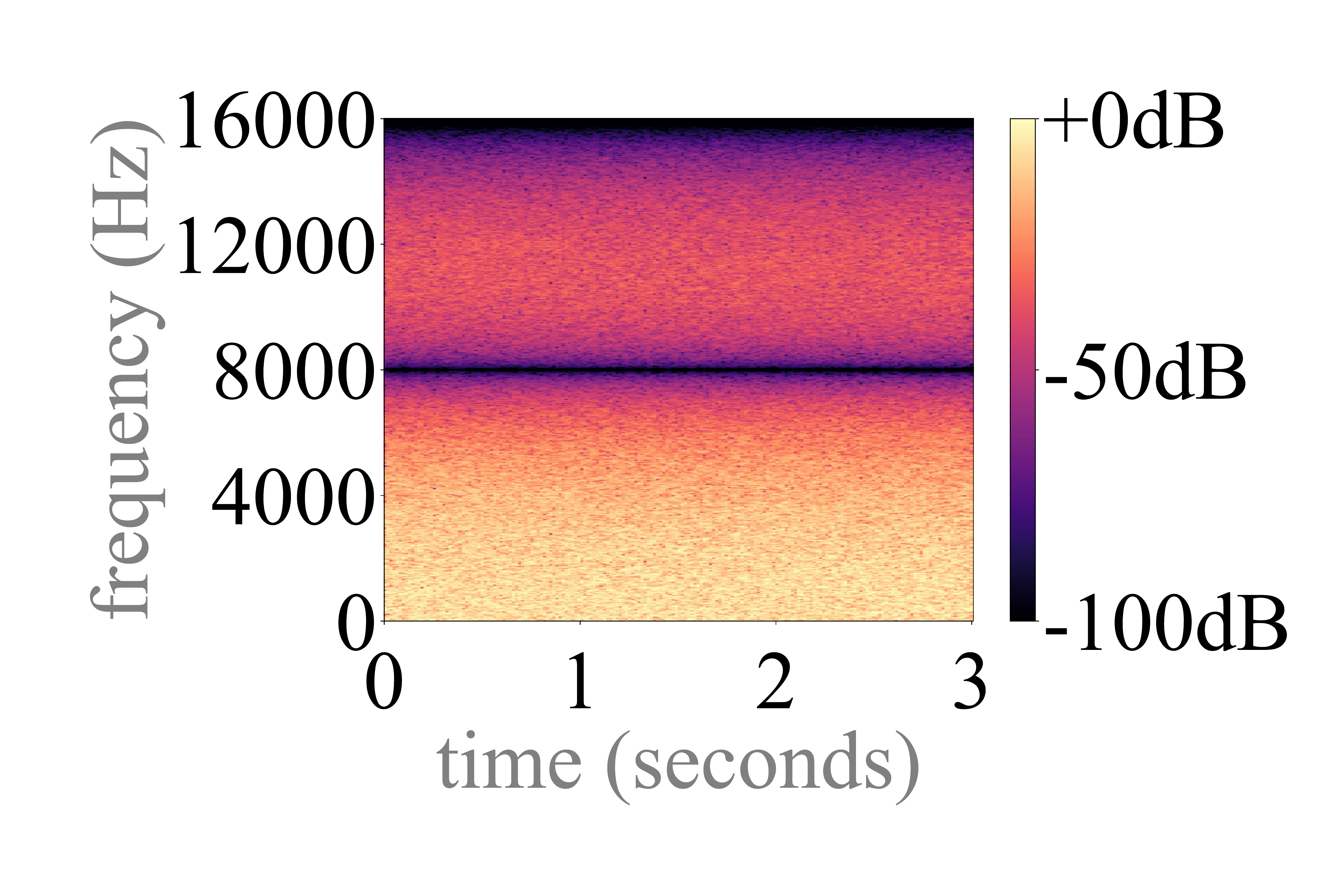}}
		\vspace{-3mm}
		{(f) Interpolation:\\linear}\medskip
	\end{minipage}
	\begin{minipage}[b]{0.18\linewidth}
		\centering
		\centerline{\includegraphics[width=1.1\linewidth]{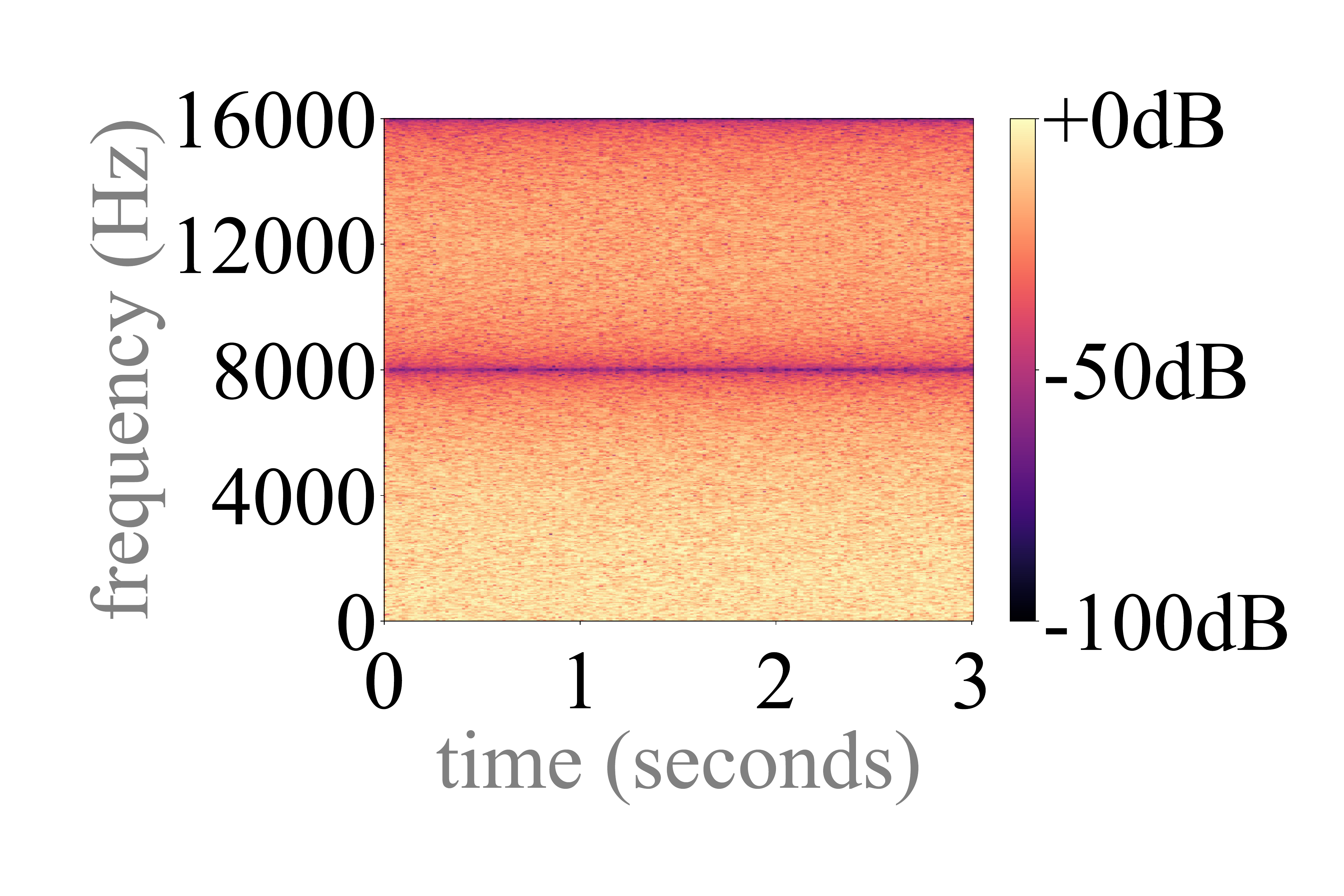}}
		\vspace{-3mm}
		{(g) Interpolation:\\nearest neighbor}\medskip
	\end{minipage}
	\begin{minipage}[b]{0.18\linewidth}
		\centering
		\centerline{\includegraphics[width=1.1\linewidth]{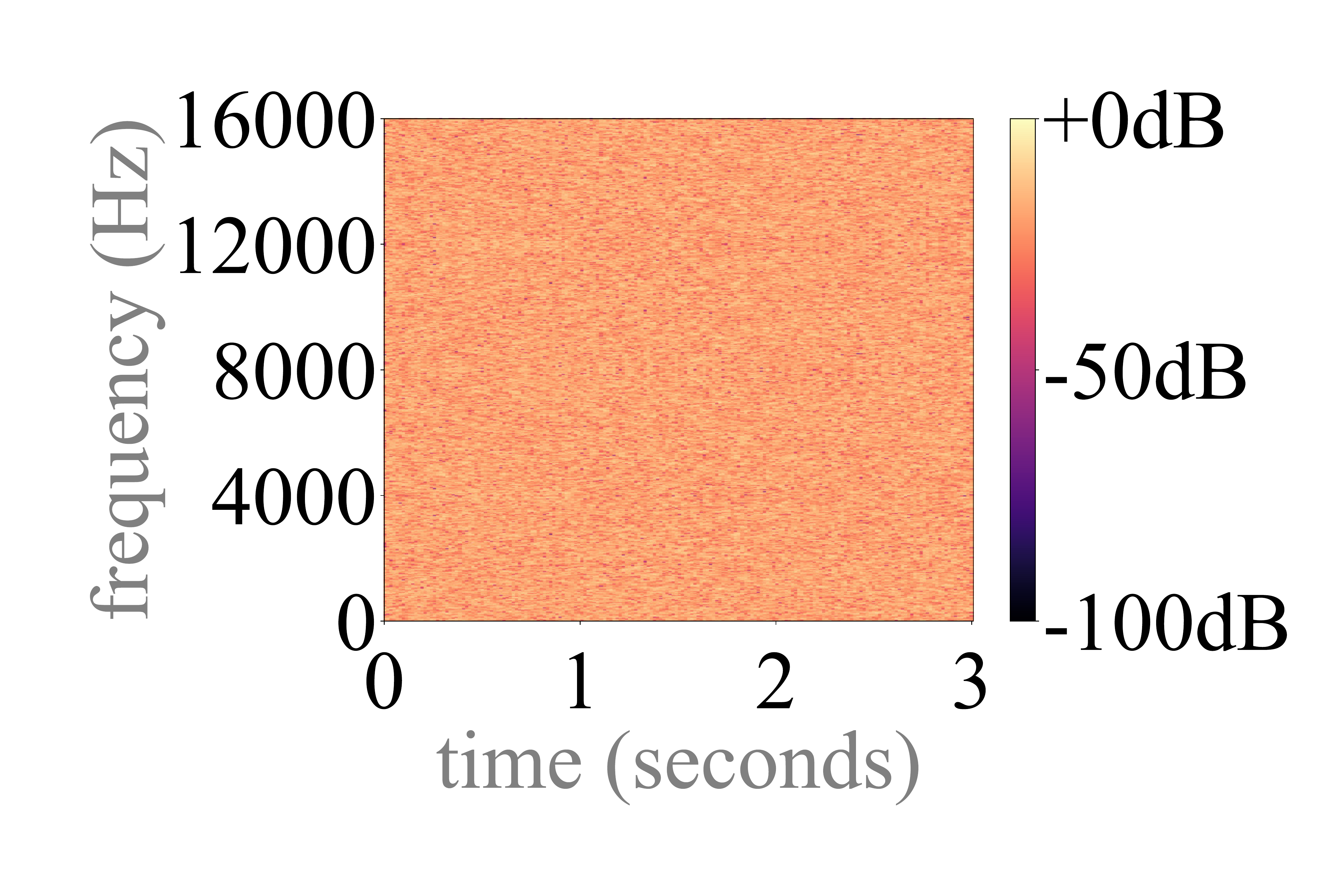}}
		\vspace{-3mm}
		{(h) Interpolation:\\stretch}\medskip
	\end{minipage}
	\vfill \vspace{-5mm}
	\caption{{Input: white noise at 8kHz. Upsampling~($\uparrow$4) layers  can introduce filtering artifacts that attenuate some bands. Only~(e, f, g) ``horizontal valleys" are considered filtering artifacts, because are caused by non-learnable interpolations. Hence,~(e, f, g) layers would introduce filtering  artifacts even after training---while the rest of ``horizontal valleys"~(a, b, c, d) can change during training. Transposed CNNs with stride=4.}}
	\vspace{-4mm}
	\label{fig:all}
\end{figure*}

\begin{figure*}[h]
			\hspace{18mm}
	\begin{minipage}[b]{0.18\linewidth}
		\centering
		\centerline{\includegraphics[width=1.1\linewidth]{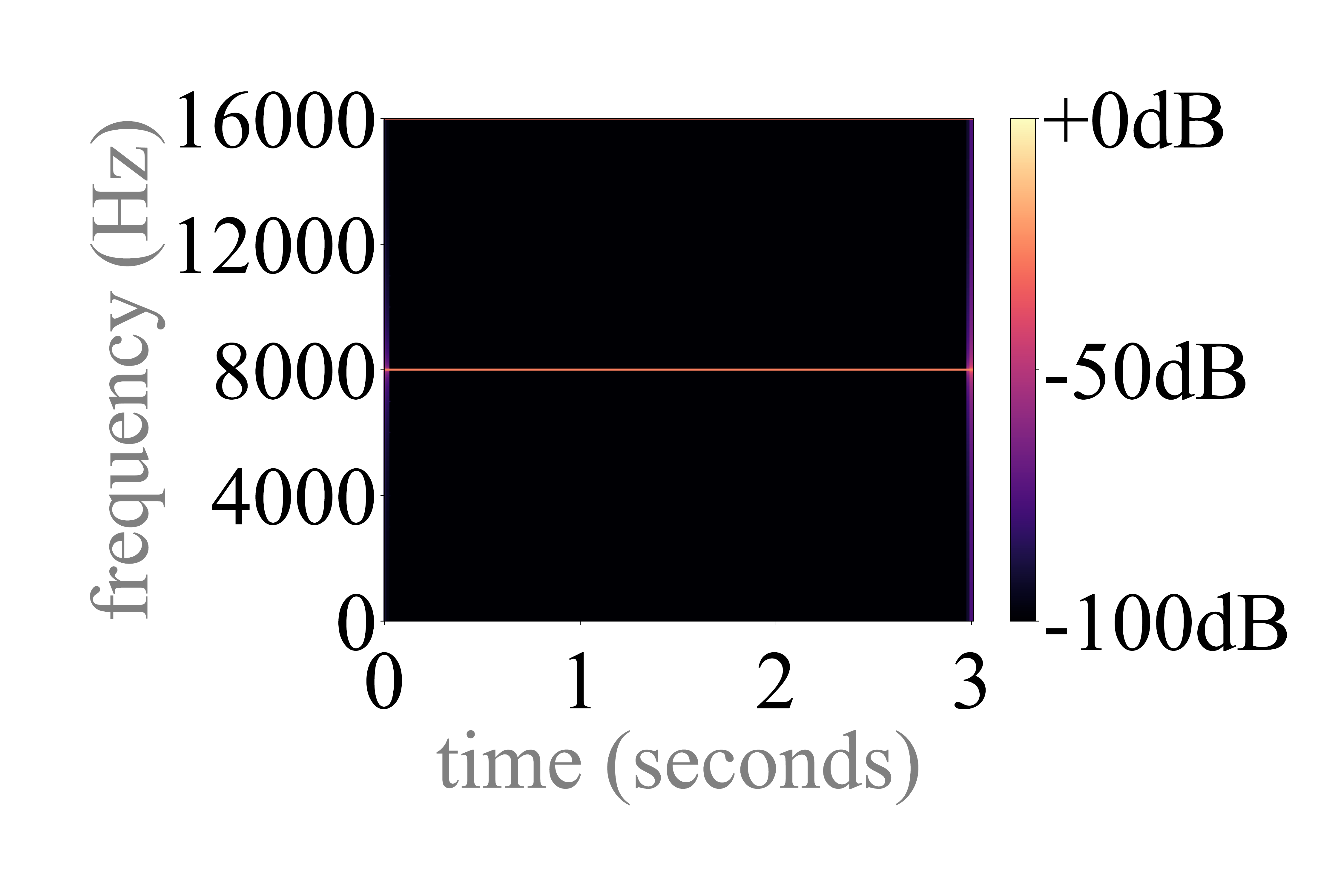}}
		\vspace{-3mm}
		{(a) Transposed CNN:\\no overlap (length=8)}\medskip
	\end{minipage}	
	\begin{minipage}[b]{0.18\linewidth}
		\centering
		\centerline{\includegraphics[width=1.1\linewidth]{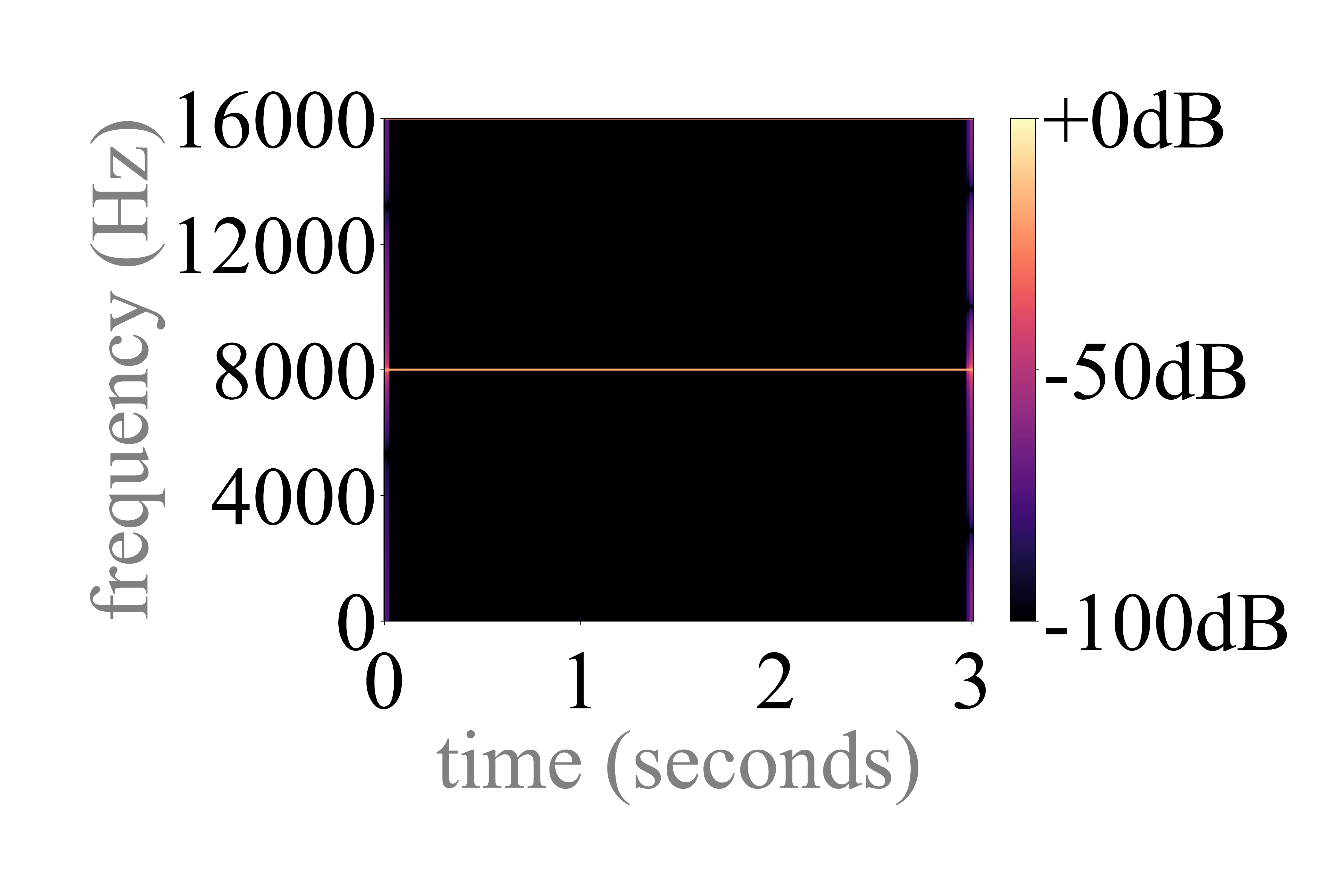}}		\vspace{-3mm}
		{(b) Transposed CNN:\\\mbox{partial overlap (length=9)}}\medskip
	\end{minipage}
	\begin{minipage}[b]{0.18\linewidth}
		\centering
		\centerline{\includegraphics[width=1.1\linewidth]{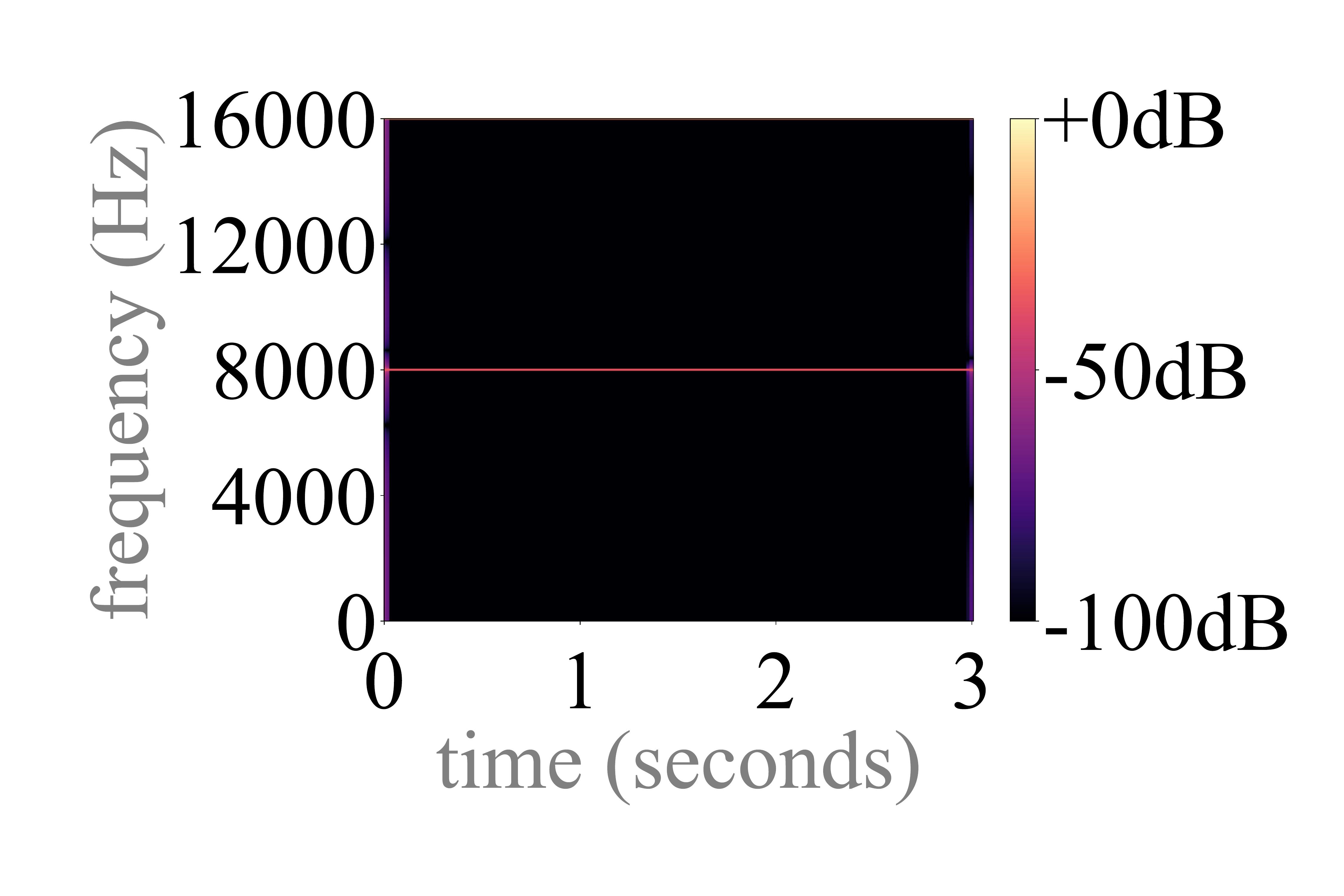}}		\vspace{-3mm}
		{(c) Transposed CNN:\\full overlap (length=4)}\medskip
	\end{minipage}
	\begin{minipage}[b]{0.18\linewidth}
		\centering
		\centerline{\includegraphics[width=1.1\linewidth]{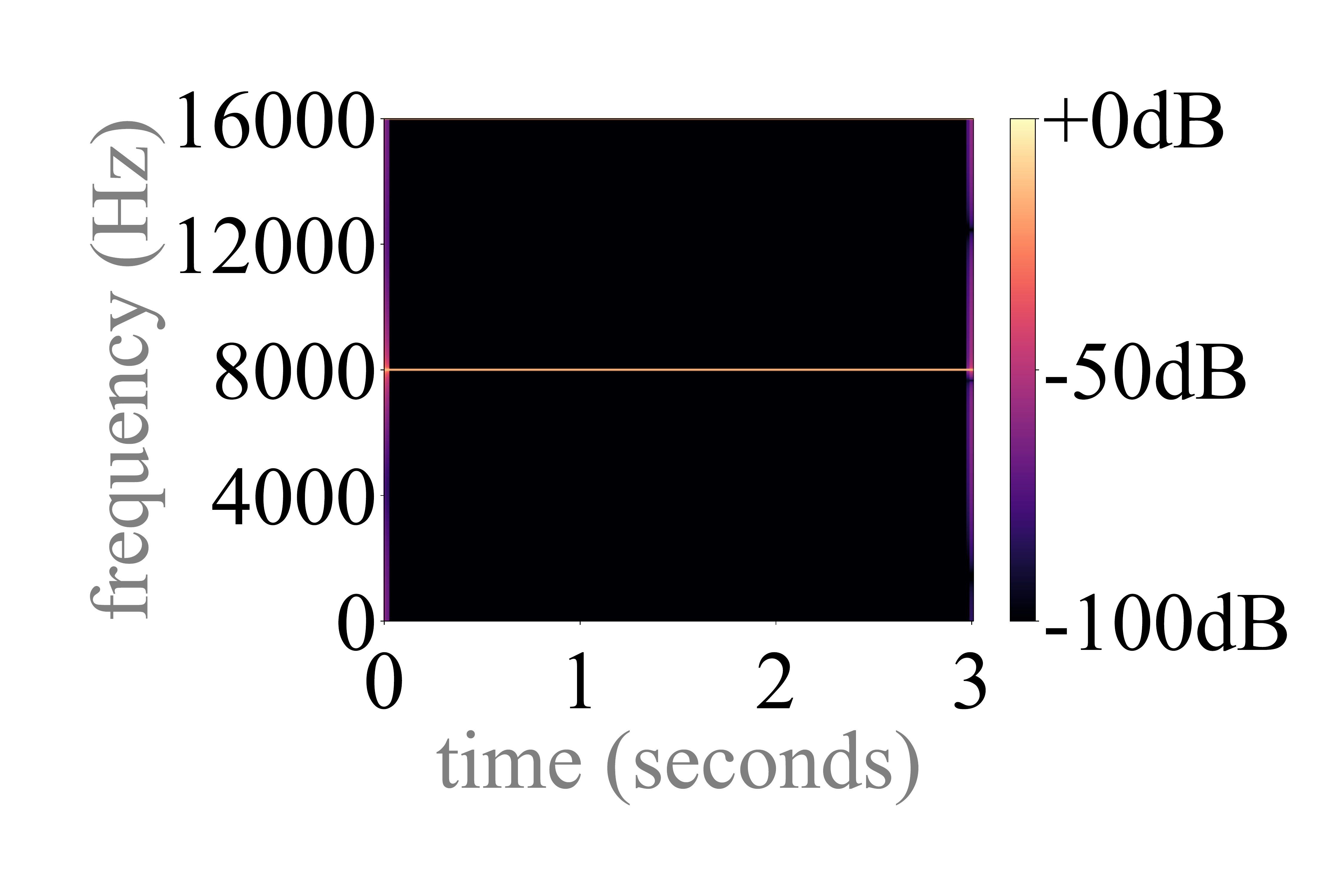}}		\vspace{-3mm}
		{(d) Subpixel CNN\\(filter length=9)}\medskip
	\end{minipage}
	\vfill \vspace{-4mm}
			\hspace{18mm}
	\begin{minipage}[b]{0.18\linewidth}
		\centering
		\centerline{\includegraphics[width=1.1\linewidth]{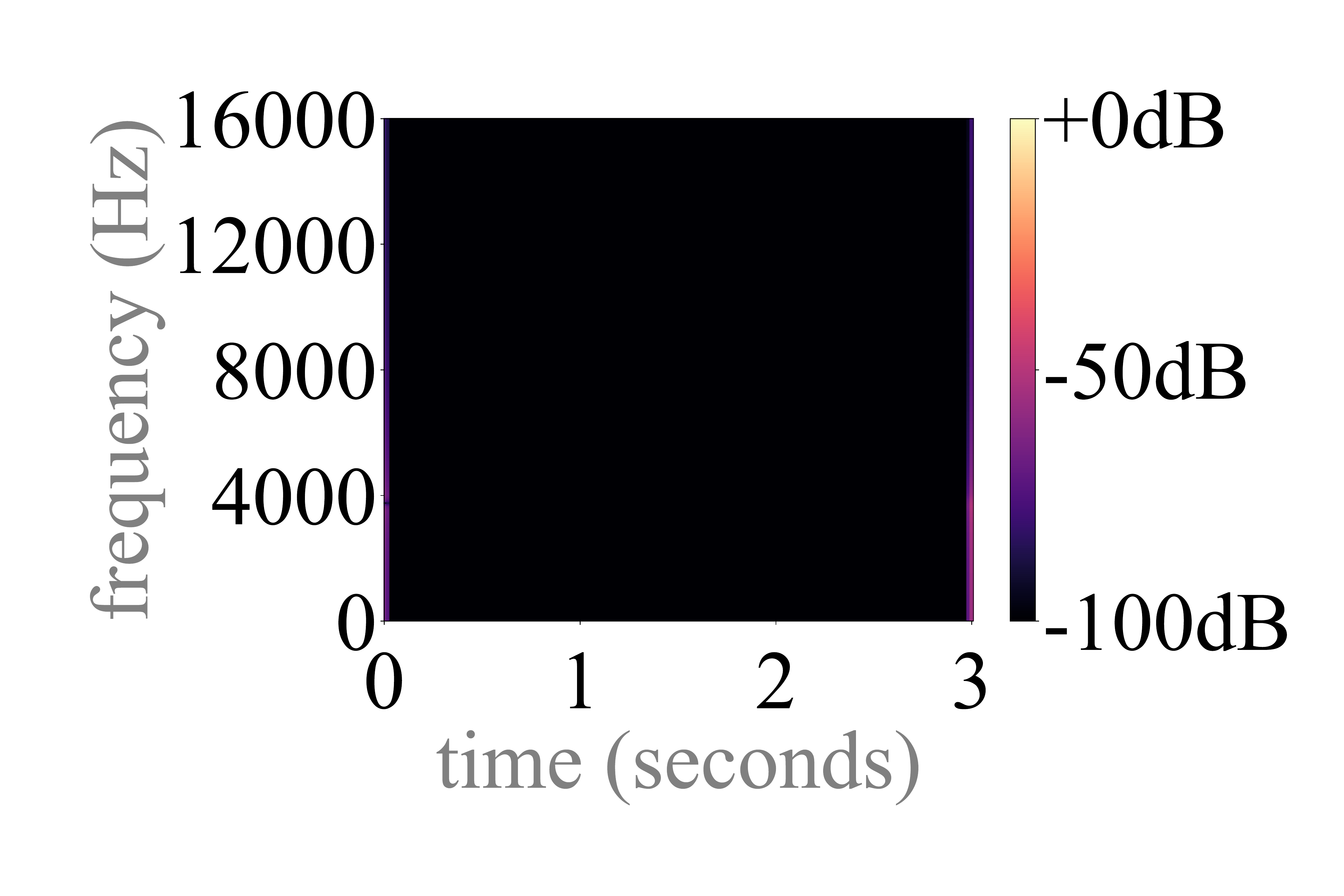}}		\vspace{-3mm}
		{(e) Interpolation:\\sinc}\medskip
	\end{minipage}	
	\begin{minipage}[b]{0.18\linewidth}
		\centering
		\centerline{\includegraphics[width=1.1\linewidth]{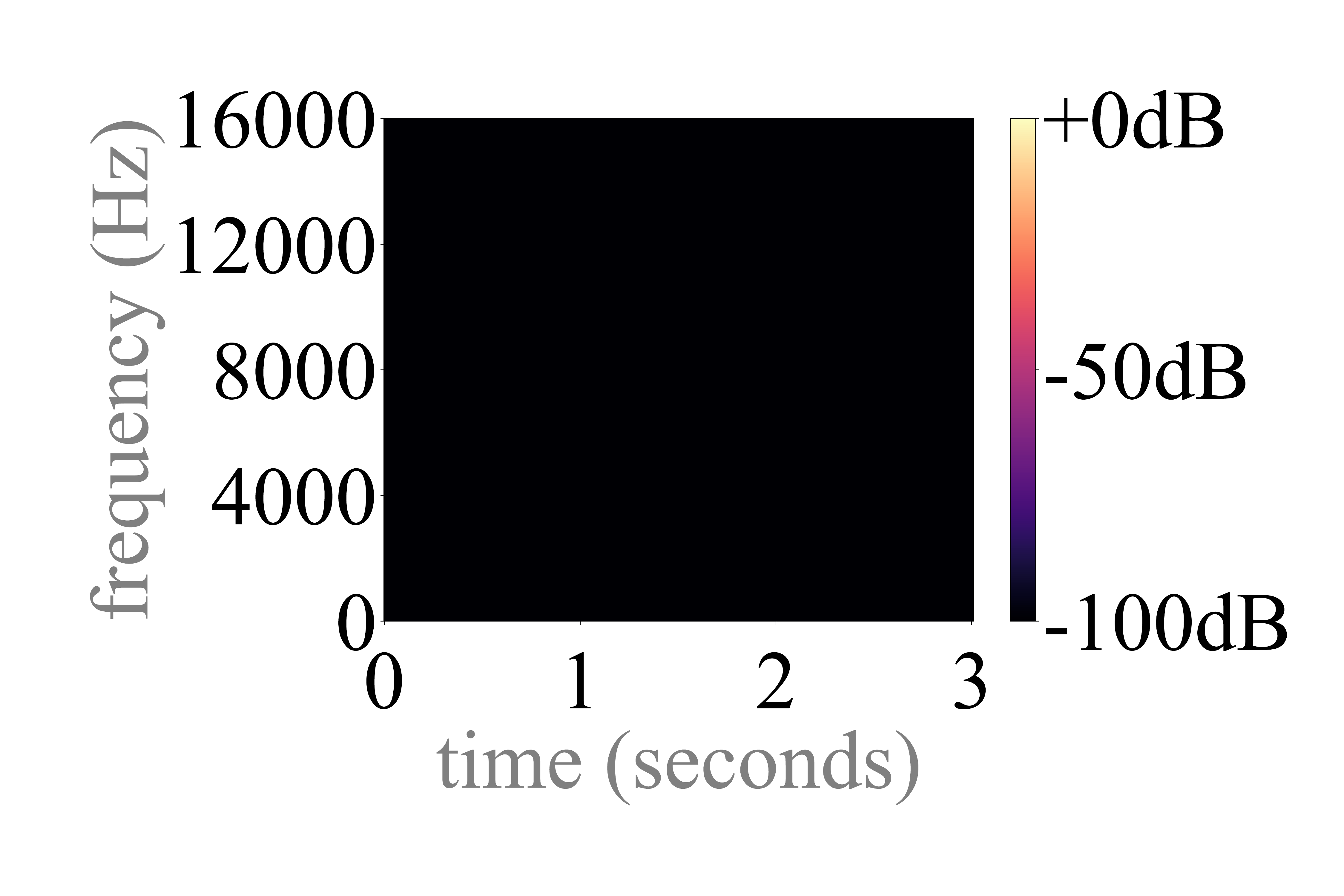}}		\vspace{-3mm}
		{(f) Interpolation:\\linear}\medskip
	\end{minipage}
	\begin{minipage}[b]{0.18\linewidth}
		\centering
		\centerline{\includegraphics[width=1.1\linewidth]{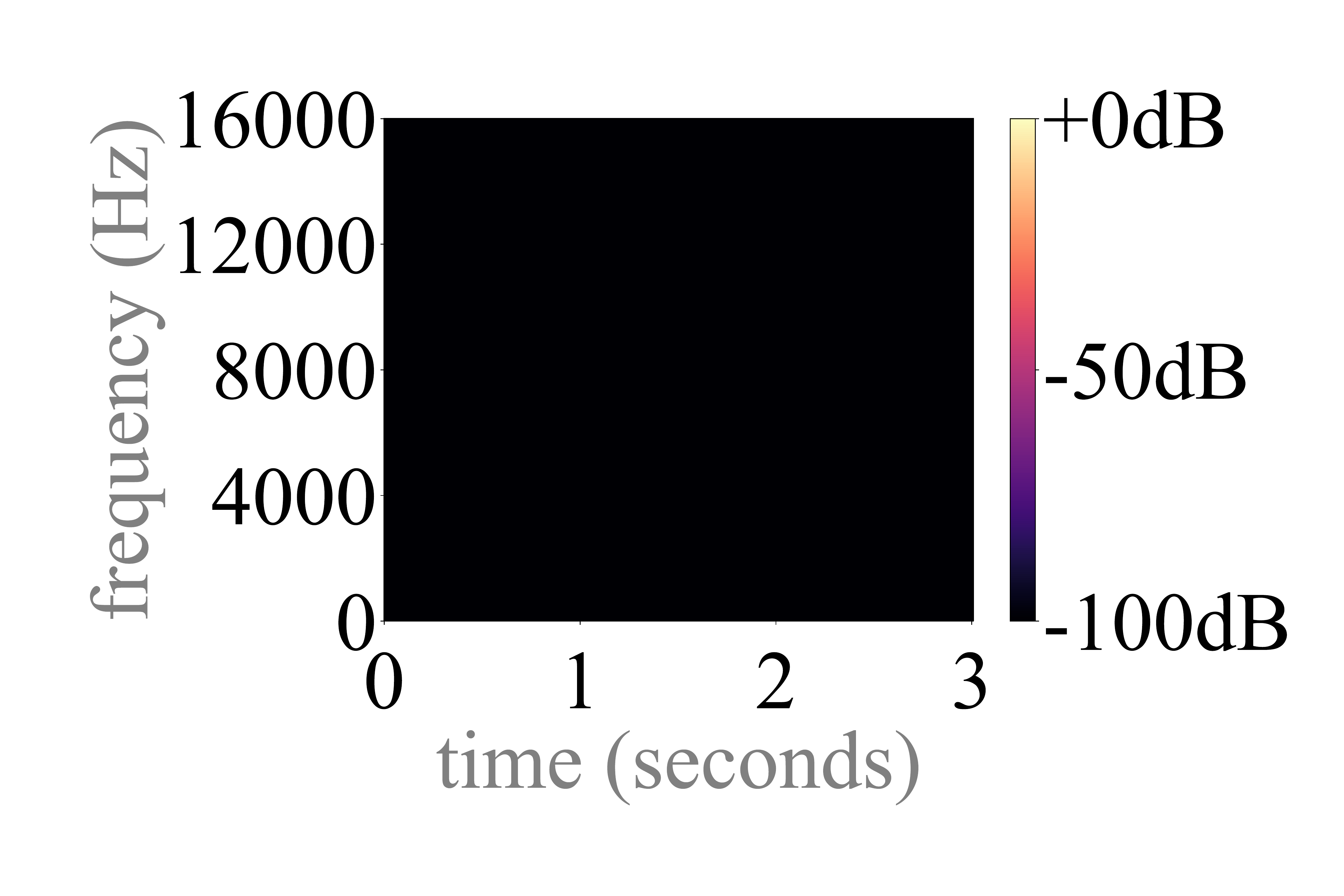}}		\vspace{-3mm}
		{(g) Interpolation:\\nearest neighbor}\medskip
	\end{minipage}
	\begin{minipage}[b]{0.18\linewidth}
		\centering
		\centerline{\includegraphics[width=1.1\linewidth]{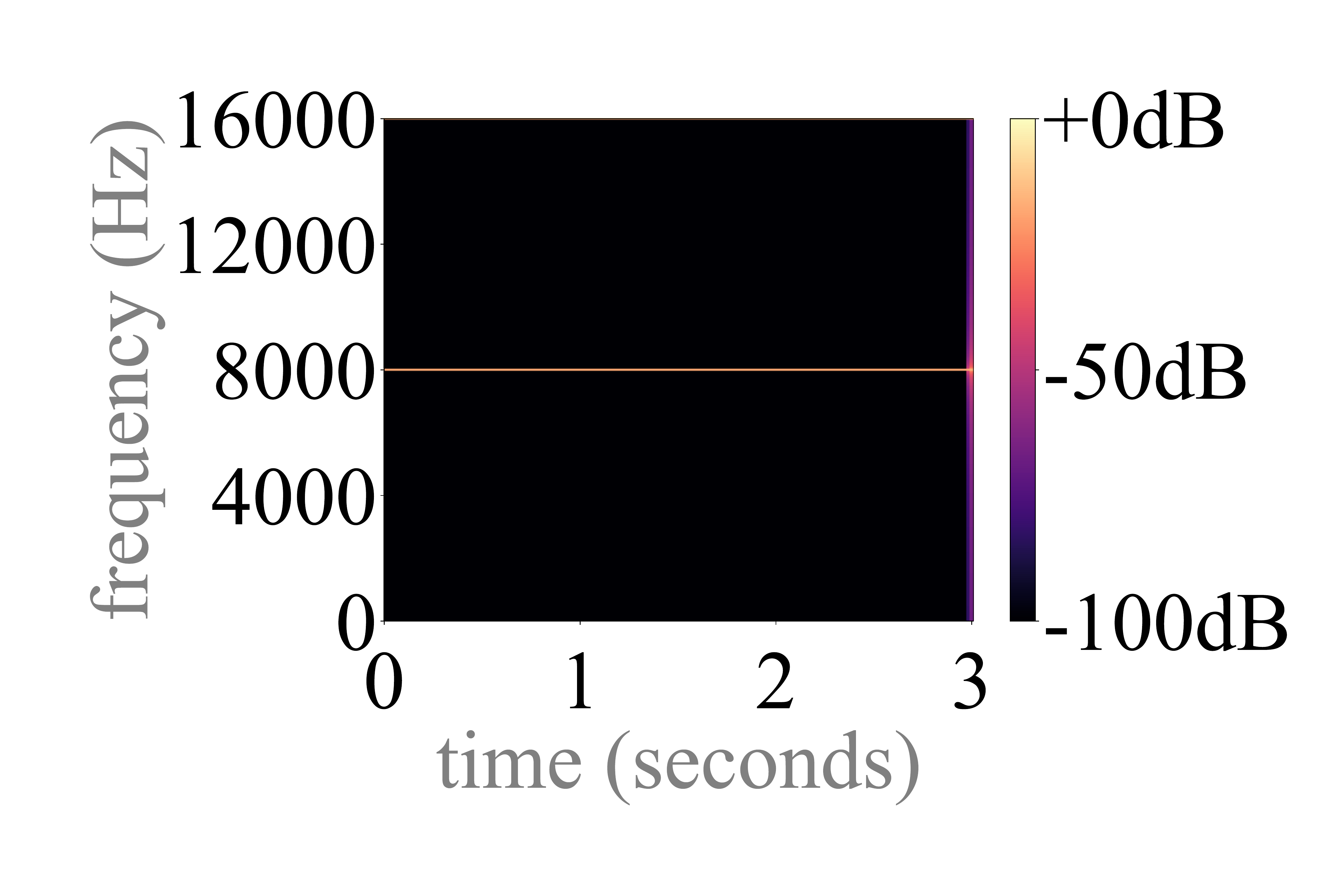}}		\vspace{-3mm}
		{(h) Interpolation:\\stretch}\medskip
	\end{minipage}
	\vfill \vspace{-5mm}
	\caption{\mbox{Input: ones~(constant) at 8kHz. Upsampling~($\uparrow$4) layers can introduce tonal artifacts~(horizontal lines).  Transposed CNNs with stride=4.}}
	\vspace{-3mm}
	\label{fig:all}
\end{figure*}

\vspace{-2mm}
\section{Upsampling layers}
\vspace{-1.5mm}
\subsection{Transposed convolution and tonal artifacts}
\vspace{-1mm}

Transpososed convolutions can introduce undesired tonal artifacts due to {(i)}~their weights' initialization, {(ii)}~the loss function, and {(iii)}~overlap \mbox{issues~\cite{pons2020upsampling,odena2016deconvolution}. 
Yet, how challenging is to mitigate them?}
%The {loss function} issue relates to the way learning is defined, while the weight's initialization and overlap issues are caused by the way the model is initialized and constructed, respectively. 
{The weights' initialization issue~{(i)} is caused because randomly initialized transposed convolutional filters repeat accross time~\cite{pons2020upsampling,kumar2019melgan}. 
This issue is commonly addressed via learning from data, since training may help mitigating the tonal artifacts caused by this problematic initialization. 
Loss function issues~{(ii)} emerge when convolutional neural networks (CNNs) are used as a loss, since it involves a transposed convolution step during backpropagation~\cite{odena2016deconvolution}. 
Loss issues can be avoided by just not using adversarial~\cite{donahue2018adversarial,kumar2019melgan,pascual2017segan} or deep feature losses~\cite{germain2019speech}. 
Finally, overlap {issues~{(iii)} can be mitigated by carefully choosing the filter-length and stride~\cite{kumar2019melgan}.} More precisely, one can omit overlap issues by using no overlap~(length$=$stride) and full overlap~({length} is multiple of the {stride}) setups, but partial overlap setups ({length} is not multiple of the stride) would still introduce tonal artifacts due to overlap issues~\cite{pons2020upsampling}. Yet, importantly, even under no overlap and full overlap setups, the loss function and the weights' initialization issues remain. One can observe that tonal artifacts emerge after \mbox{(random) initialization in Fig.~2~(a, b, c).}

\vspace{-3mm}
\subsection{Interpolation upsamplers and filtering artifacts}
\vspace{-1mm}

Interpolation upsamplers first interpolate a given feature map and then employ a learnable convolutional layer afterwards~\cite{odena2016deconvolution}. They do not cause tonal artifacts, but can introduce filtering artifacts~\cite{pons2020upsampling}. 
Such artifacts attenuate some bands, and are introduced by the~(fixed, non-learnable) frecuency response of the interpolation, see \mbox{Fig.~1~(e, f, g).}
Filtering artifacts vary depending on the frequency response of the interpolation operator: % We study the following ones:

\noindent-- \emph{Stretch interpolation} upsamples the signal with zeros.  
%\vspace{-3mm}
%\[x_M[n]=\sum_{k=-\infty}^{\infty} x[k] \delta[n-kM],\vspace{-3mm}\]
%\noindent where $M$ is the upsampling factor. I
Its frequency response is:
%\vspace{-1mm}
$X_{\text{stretch by} M}({\rm e}^{j\omega}) = X({\rm e}^{j\omega M})$, where $M$ is the upsampling factor.
% https://www.youtube.com/watch?v=kKb_9rmTnHY&feature=emb_title
Hence, stretch interpolation scales the frequency axis by $M$, exposing the spectral replicas without further transforming the signal. In other words: its flat frequency response does not introduce filtering artifacts, see Fig.~1~(h).

%Hence, stretch interpolation scales the frequency axis by $M$, exposing spectral replicas. Its flat frequency response does not color the signal, what does not introduce filtering artifacts, see Figure 1~(j).
%It has a flat frequency response that does not color the signal~(without filtering artifacts), see Figure 1~(j).

\noindent-- \emph{Sinc interpolation} can be implemented as a stretch interpolation~+~convolution with a sinc filter~\cite{smith2007mathematics}. 
It is known as bandlimited interpolation since the frequency response of a sinc is a low-pass filter that removes the introduced spectral replicas, see Fig.~1~(e).

\noindent-- \emph{Nearest neighbor interpolation}: can be implemented as stretch interpolation~+~convolution with a rectangular filter~\cite{pons2020upsampling}. Since the frequency response of a rectangular filter is a sinc, its filtering artifacts attenuate~(some) high-frequency bands, see Fig.~1~(g).
%The frequency response of a rectangular filter is a sinc($\cdot$). Accordingly, its filtering artifacts attenuate~(some) high-frequency bands, see Fig.~1~(g).

\noindent-- \emph{Linear interpolation} can be implemented as stretch interpolation~+~convolution with a triangular filter~\cite{pons2020upsampling}. Since the frequency response of a triangular filter is a sinc$^2$, 
%The frequency response of a triangular filter is a sinc$^2$($\cdot$).
%Accordingly,
its filtering artifacts attenuate the high-frequency bands---that are more attenuated than when using nearest {neighbor interpolation, see Fig.~1~(f, g).}

%\noindent-- \emph{Learnable interpolation}. Since many interpolations can be implemented as stretch interpolation + convolution, we experiment with~(non pre-defined) learnable convolutions.\footnote{Interpolation-based upsamplers are conformed by interpolation + learnable convolution. The interpolation step can be implemented as stretch interpolation + convolution. Hence, learnable interpolations are: stretch interpolation + learnable convolution + learnable convolution.}~Its filtering artifacts, then, depend on the learnt frequency response.

%\noindent-- \emph{Learnable interpolation}. Since many interpolations can be implemented as stretch interpolation + convolution, we experiment with~(non pre-defined) learnable convolutions.~Its filtering artifacts, then, depend on the learnt frequency response.

\vspace{1mm}

\noindent Remember that the non-learnable interpolations we have just discussed (e.g.:~sinc or linear, explained as stretch interpolation + non-learnable convolution) are typically followed by a learnable convolution.
To the best of our knowledge, we are the first to experiment with stretch and sinc interpolation layers for neural audio synthesis.%Provided that we have introduced non-learnable interpolations as stretch interpolation + non-learnable convolution, interpolation upsamplers can be regarded as: stretch interpolation + non-learnable convolution + learnable convolution.

\begin{figure}[t]
	\vspace{2mm}
	\centering
	\centerline{\includegraphics[width=\columnwidth]{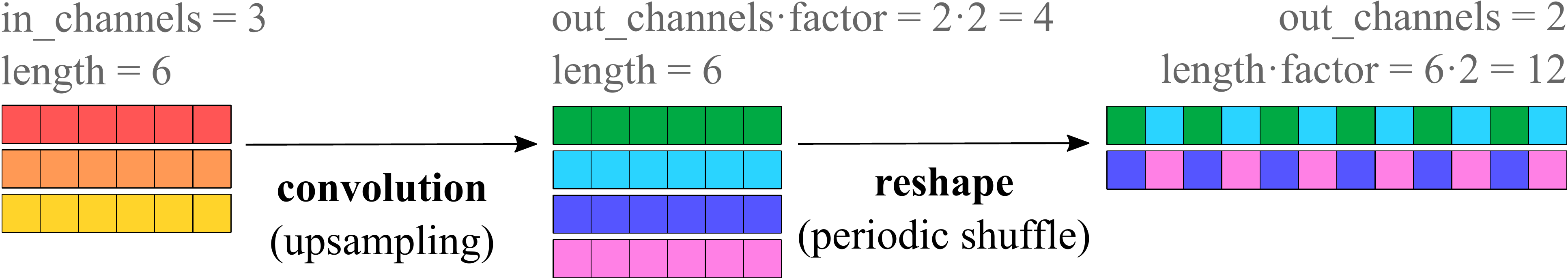}}
	\vspace{-3mm}	
	\caption{Subpixel convolution: reshaping can introduce tonal artifacts.}
	\vspace{-3mm}
\end{figure}

\vspace{-2mm}

\subsection{Subpixel convolution and  tonal artifacts}
\vspace{-1mm}

Subpixel convolution is based on convolution and reshape~\cite{pons2020upsampling,pandey2020densely,shi2016real}. The convolution upsamples the signal along the channel axis, and the reshape operation
is based on periodic shuffling---that reorders the convolution's output to match the desired output shape, see Fig.~3. 
It can introduce tonal artifacts due to the periodic shuffle operator, since it interleaves
consecutive samples out of convolutional filters having different
weights 
(depicted in Fig.~3 as colored periodicities).
Such periodicities are caused by the~(different) dynamics each feature map can exhibit, since interleaving activations with different energies leads to periodic~(tonal) artifacts~\cite{aitken2017checkerboard}.
This can be particularly noticeable after initialization, since feature maps can exhibit different dynamics due to the random weights of the filters as in Fig.~2~(d).

	\vspace{-3mm}

\subsection{Wavelet-based upsamplers and filtering artifacts}
	\vspace{-1mm}

Cascaded filter banks are often used to implement multi-resolution discrete wavelet transforms~(Fig.~4) \cite{sweldens1998lifting}.
One interesting property of wavelets is that they allow for perfect reconstruction. Thus meaning that one can recover the input signal after following the analysis/synthesis steps. Hence, no additive~(tonal) or substractive~(filtering) artifacts emerge after downsampling/upsampling with wavelets. 
%Yet, wavelets are not tuned to perform any task, because these are designed to perfectly reconstruct the input. 
However, the perfect reconstruction property only holds in absence of any processing in the wavelet domain---and we describe wavelet-inspired downsampling/upsampling layers that, combined with neural networks, have the capacity to perform downstream tasks~\cite{nakamura2020time,sweldens1998lifting}:

\noindent-- \emph{Lazy wavelet layers} downsample the signal into odd/even samples, and interleave odd/even samples for upsampling.

\noindent-- \emph{Haar wavelet layers} can be described following the filter bank scheme in Fig.~4, where analysis/synthesis filters are set as: 
\vspace{-2mm}
\[L_a[n] = [\frac{1}{\sqrt{2}}, \frac{1}{\sqrt{2}}], \hspace{4mm} H_a[n] = [\frac{1}{\sqrt{2}}, -\frac{1}{\sqrt{2}}]\]
\vspace{-4mm}
\[L_s[n]=L_a[-n],  \hspace{6mm}  H_s[n]=H_a[-n]\]

\noindent
Importantly, the above wavelets' upsampling paths have an overall frequency response that is flat. Note this in {Fig.~5~(b, d),} were we depict that Haar wavelets' upsampling paths~(after summing all upsampling filters' outputs) have a flat overall frequency response, allowing the above wavelets to compensate for filtering artifacts.
To further undestand this idea, it is illustrative to compare the frequency response of the nearest neigbor \mbox{layer~(Fig.~5: a, c)}, and the low-pass filter of Haar wavelets'~(Fig.~5: b, d) to note their resemblance. 
Hence, the alternative signal paths in wavelets allow compensating the filtering artifacts typically introduced by, e.g., nearest neighbor.

Further, wavelets provide a principled way to downsample.~Inspired by that, Nakamura and Saruwatari~\cite{nakamura2020time} replaced \mbox{WaveUnet's~\cite{stoller2018wave}} downsampling~(discarding every other time step) and upsampling {(linear interpolation) layers by lazy and Haar wavelet layers. How}ever, the above wavelet layers are designed to downsample/upsample x2, as the original WaveUnet. In our work, we extend those to downsample/upsample~x4 via cascading wavelets as in Fig.~4.

We also experiment with \emph{learnable wavelet layers}. To do so, we implement the above wavelets~(both analysis and synthesis) using the lifting scheme~\cite{nakamura2020time,sweldens1998lifting} that defines wavelet transforms with 3 parameters: a prediction operator $P$, an update operator $U$, and a normalization constant $A$~\cite{nakamura2020time,sweldens1998lifting}. Haar and lazy wavelets can be implemented using the lifting scheme by setting $P$$=$$1$, $U$$=$$0.5$, $A$$=$$\sqrt{2}$; and $P$$=$$0$, $U$$=$$0$, $A$$=$$1$~(respectively)~\cite{nakamura2020time,sweldens1998lifting}.
By relying on the lifting scheme, wavelet layers can be easily implemented in a differentiable fashion where $P$, $U$ and $A$ can be learnable parameters---what allows to easily define {learnable wavelet downsampling/upsampling layers}. For our learnable wavelet experiments, we initialize such parameters to $P$$=$$0$, $U$$=$$0$, $A$$=$$1$~(as the lazy wavelet, since it gave better results).
We are the first to study learnable wavelet layers for audio synthesis.

% https://en.wikipedia.org/wiki/Generalized_lifting

\begin{figure}[t]
	\vspace{-7mm}
	\centering
	\centerline{\includegraphics[width=1.0\columnwidth]{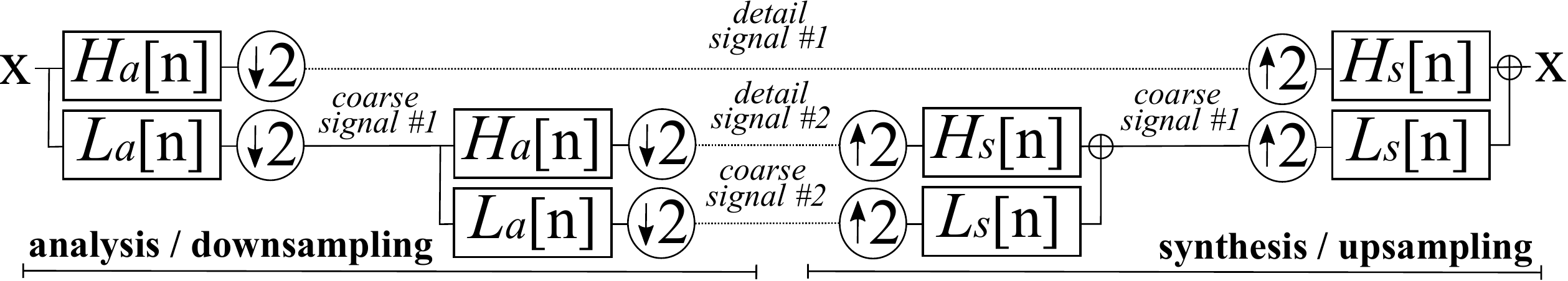}}
	\vspace{-2mm}	
	\caption{Cascaded wavelets: with low-pass $L_a[n]$, $L_s[n]$ and high-pass $H_a[n]$, $H_s[n]$ filters. $\downarrow$2 operator discards every other time step. $\uparrow$2 operator refers to stretch interpolation, upsampling with zeros.}
	\vspace{-4mm}
\end{figure}

	\vspace{-3mm}
\subsection{Discussion: pros and cons of each upsampling layer}
	\vspace{-1mm}

So far, we discussed filtering and tonal artifacts. However, spectral replicas 
can introduce additional upsampling artifacts. In this section, we consider them to further study the above upsampling layers:

\noindent -- \emph{Spectral replicas of signal offsets}. Offsets are constants with zero frequency. Hence, its frequency transform contains an energy
component at frequency zero. When upsampling, zero-frequency spectral replicas can appear in-band, introducing tonal artifacts. Specifically, they can appear at multiples of ``sampling rate / upsampling factor" Hz~(the sampling rate being the one of the upsampled signal). For example, in Fig.~2 such replicas appear at multiples of 32kHz / 4 = 8kHz.
To further understand this concept, note that stretch interpolation does not introduce tonal artifacts. Yet, {Fig.~2~(h)} depicts tonal artifacts at multiples of 8kHz. This is because the spectral replicas of the ones~(constant) signal appear in-band when upsampling~\cite{pons2020upsampling}.

\noindent -- \emph{Spectral replicas of offsets interact with filtering artifacts}. Note that interpolation-based upsamplers~(which introduce filtering artifacts) can attenuate the exact bands where spectral replicas of signal offests appear. Namely, they can attenuate around ``sampling rate / upsampling factor" Hz as depicted in Figs. 1~(e, f, g) and 5. Hence, filtering artifacts are a powerful tool to combat the spectral replicas of signal offsets. Note that no tonal artifacts due to signal offsets appear when uspampling a ones~(constant) signal in Fig.~2~(e, f, g).

\noindent -- \emph{Spectral replicas of offsets interact with tonal artifacts}. The tonal artifacts introduced by transposed and subpixel convolutions occur at the same frequency as the first  signal offset spectral replica~(at ``sampling rate / upsampling factor" Hz). \mbox{Note in Fig.~2~(a, b, c, d)} that tonal artifacts appear at 8kHz~(due to transposed and subpixel convolutions, and signal offsets) and at 16kHz~(due to signal offsets). Hence, which introduces stronger artifacts? Transposed~and~subpixel convolutions, or the spectral replicas? \mbox{We now address this question.}

\noindent -- \emph{The relative energy of tonal artifacts}. 
Remarkably, we do not observe tonal artifacts in Fig.~1. Hence, the tonal artifacts introduced by transposed and subpixel convolutions~(prone to introduce tonal artifacts) have little energy, compared to the energy of the signal being upsampled (zero-mean white noise). In contrast, Fig.~2 depicts tonal arifacts with similar energy for:~(i) transposed and subpixel convolutions (prone to introduce tonal artifacts), and~(ii) stretch interpolation (which does not introduce tonal artifacts). These observations denote that  the spectral replicas of signal offsets can introduce stronger tonal artifacts than transposed and subpixel convolutions. 
For this reason, in section 3, we include two modifications meant to reduce signal offsets accross feature maps: {removing learnable bias terms in our models,}  and studying to incorporate normalization layers.

\noindent -- \emph{Spectral replicas as a source of high-frequency content}.~We~can sort interpolation upsamplers by how strong their filtering artifacts are: {sinc$\to$linear$\to$nearest neighbor$\to$stretch, see Fig.~1}. Note that sinc interpolation strongly filters the signal, and stretch interpolation introduces no filtering artifacts.
While sinc interpolation is widely used in audio because it removes all spectral replicas, this might not be desirable for deep learning. We hypotesize that allowing spectral replicas accross feature maps is beneficial, as it allows the model to have access to coherent high-frequency feature maps for wide-band synthesis. 
In section 3 we experimentally validate this hypothesis.

\begin{figure}[t]
	\vspace{-8.5mm}
	\begin{minipage}[b]{0.49\columnwidth}
		\centering
		\centerline{\includegraphics[width=0.7\columnwidth]{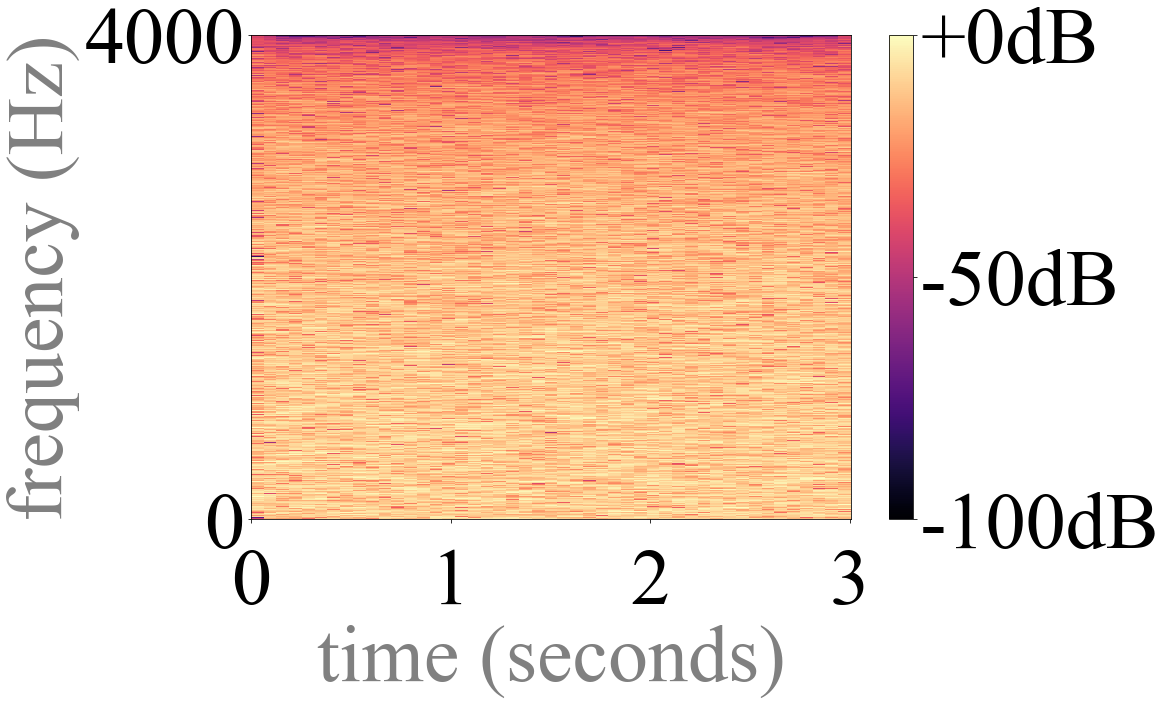}}
				\vspace{-0.5mm}
		{(a) Nearest neighbor $\uparrow$2}\medskip
	\end{minipage}
\vspace{-1mm}
	\begin{minipage}[b]{0.49\columnwidth}
		\centering
		\centerline{\includegraphics[width=0.61\columnwidth]{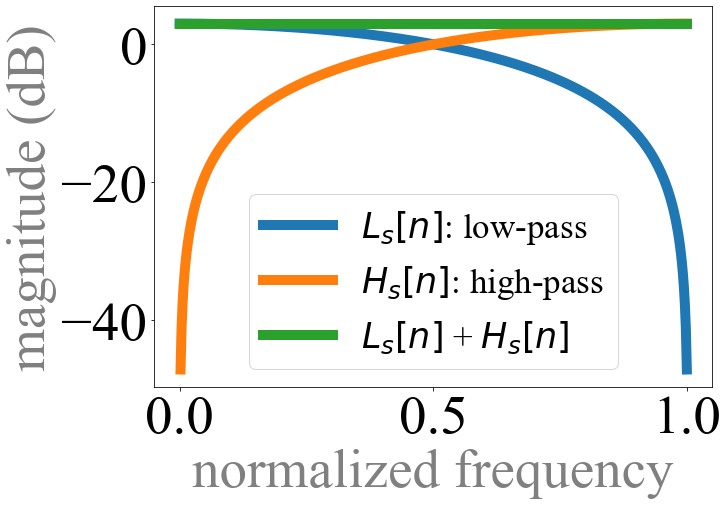}}
		\vspace{-0.5mm}
		{(b) Haar wavelet $\uparrow$2}\medskip
	\end{minipage}	
	\begin{minipage}[b]{0.49\columnwidth}
		\centering
		\centerline{\includegraphics[width=0.7\columnwidth]{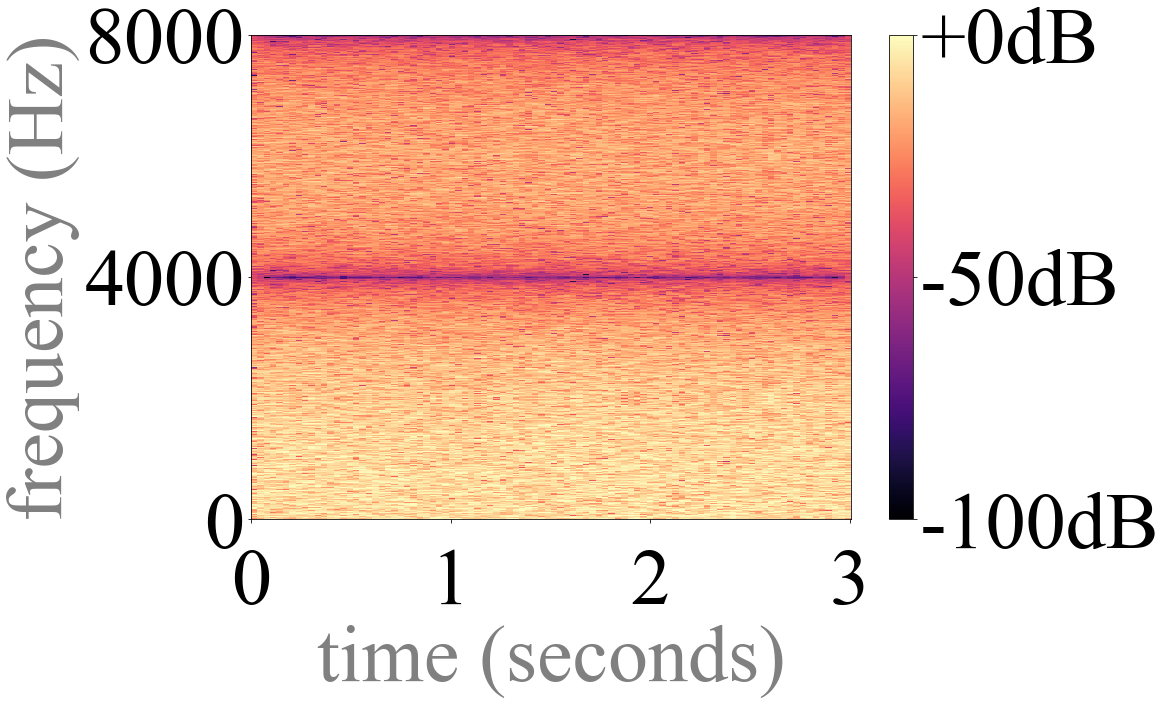}}
				\vspace{-0.5mm}
		{(c) Nearest neighbor $\uparrow$4}\medskip
	\end{minipage}
	\begin{minipage}[b]{0.49\columnwidth}
		\centering
		\centerline{\includegraphics[width=0.61\columnwidth]{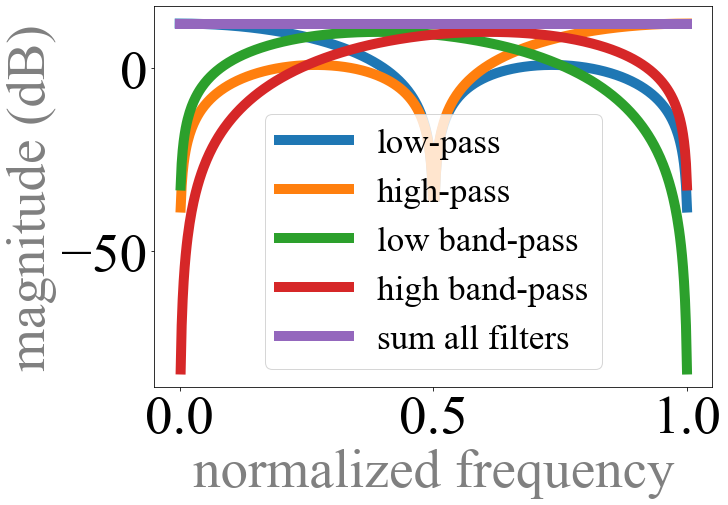}}
				\vspace{-0.5mm}
		\mbox{(d) Cascaded Haar wavelets $\uparrow$4}\medskip
	\end{minipage}
	
	\vspace{-5mm}
	\caption{{The frequency response of the nearest neighbor upsampler corresponds to the low-pass synthesis filter of the depicted wavelets.} Left: white noise at 4kHz upsampled~($\uparrow$) by 2 or 4. Right: frequency response of the synthesis filters of Haar wavelet layers.}
	\vspace{-4mm}	
\end{figure}

\vspace{-2mm}
\section{experimenting with upsampling layers}
\vspace{-1.25mm}
 
We use the MUSDB~\cite{musdb18} music source separation benchmark to experiment with the above upsampling layers. It is composed of 150 stereo songs~(86 train, 14 validation, 50 test) at 44.1kHz. For each song, 4 stereo sources are extracted: vocals, bass, drums, and other. Table~1 reports our results on the test set, based on the average signal-to-distortion ratio~(SDR) accross all sources~\cite{vincent2006performance}.\footnote{Following previous works~\cite{stoller2018wave,defossez2019music}: for every source, we report the median over all tracks of the median SDR over each test track of MUSDB~\cite{musdb18}.} 
We compare Demucs-like models, which are modified to accomodate the upsampling layers under discussion. Our base model is the original Demucs~\cite{defossez2019music}, conformed by 6 encoding blocks~(with strided CNN, ReLU, GLU) and 6 decoding blocks~(with GLU, transposed CNN, ReLU), with skip connections and x2 LSTMs in the bottleneck~(with 3200 units each). Strided and transposed convolution layers have 100, 200, 400, 800, 1600 and 3200 filters, respectively.
We use no bias terms and the 1$^\text{st}$ encoder block has no ReLUs---since this helps reducing offsets accross feature maps, taming tonal artifacts after initialization \mbox{(see section 2.5 or~\cite{pons2020upsampling}).} Our GLU non-linearities~\cite{dauphin2017language} rely on CNN filters of length 3.
Like the original Demucs, we use: very large models, of $\approx$700M parameters; weight rescaling, so that input and output signals are of the same magnitude after initialization; and their data augmentation scheme, creating new mixes on-the-fly~\cite{uhlich2017improving}.
Further, we study two strategies to mitigate upsampling artifacts:~(i)~employing post-processing networks, as an ``a posteriori" mechanism to palliate upsampling artifacts~\cite{donahue2018adversarial,dhariwal2020jukebox}; and~(ii)~using normalization layers, as a way to reduce the spectral replicas of signal offsets. %~(see section 2.5). 
%---that we find to be a strong source of additional tonal artifacts 
Post-networks~(i) are conformed by 7 residual CNN layers~(each with 8 filters of length~7,  what is $\approx$6k parameters), and are trained keeping the ``pre-network" pre-trained and frozen. We experimented with finetuning and with bigger/deeper post-networks with equivalent results.
We use normalization layers~(ii) both for the encoder and the decoder,  
to compensate for any signal offset (also coming from skip-connections) before upsampling.
% the encoder ones can compensate signal offsets coming from skip-connections, and the decoder ones can compensate signal offsets before upsampling. 
These are instance-norm \cite{ulyanov2016instance} based, and implemented after each GLU (other normalizations performed worse). 
We  train minimizing the L1 loss with Adam~\cite{kingma2014adam} for 600 epochs at a learning rate of 0.0002 (halved every 100 epochs) with 4 v100 GPUs, using batches of 32. Memory intensive runs, \mbox{like WaveUnet and WaveletUnet ones,  use batches of 16.}

 \vspace{-3mm}
\subsection{Demucs: models based on transposed convolution}
\vspace{-1mm}

{Demucs models use strided convolutions for downsampling x4,} transposed convolutions for upsampling x4, and are set to use the same filter length and stride for both downsampling and upsampling. We study three settings: partial overlap ~(length=9, stride=4), full overlap~(length=8, stride=4), and no overlap~(length=stride=4). Table 1 shows that partial overlap underperforms its counterparts, arguably because it is more prone to tonal artifacts due to the weights' initialization and overlap issues---note that no and full overlap only need to overcome the weights' initialization issue~(see section 2.1).
Given that the no overlap variant is faster to train~(see epoch times in Table 1) and obtains better results than full overlap, we select it for a listening test. Finally, when we extend no overlap with post-networks or with normalization, we observe no improvement.

\vspace{-3mm}

\subsection{WaveShuffle: models based on subpixel convolution}
\vspace{-1mm}

WaveShuffle models use strided convolutions~(length=9, stride=4) for downsampling x4, and subpixel convolutions~(length=9, stride=1) for upsampling x4.
They achieve the best SDR results,  %, differently from the rest of the models we study, 
specially when including normalization layers. %Hence, we s
%and including normalization layers improves SDR scores.  
Hence, we select the normalized variant for a listening test. 
Yet, informal listening\footref{note1} reveals that normalization layers did not remove the tonal artifacts as intended.

\vspace{-3mm}
\subsection{WaveUnet: interpolation-based models}
\vspace{-1mm}

WaveUnet models use strided convolutions~(length=9, stride=4) for downsampling x4, and we study various interpolations for upsampling x4: sinc, linear, nearest neighbor, and stretch~(see section~2.2). In section 2.5, we hypothesized that allowing spectral replicas accross feature maps can be beneficial. Here, we confirm that stretch and nearest neigbor~(allowing high-frequency spectral replicas) obtain better SDR scores than sinc and linear~(attenuating high-end frequencies).
Post-networks and normalization do not improve results drastically, for this reason we select the base stretch and nearest neigbor models for a listening test. 
Finally, the WaveUnets we consider obtain significantly better results that the original ones (5dB vs. 3dB SDR), mostly thanks to an increase in model size (from 10M to \mbox{700M learnable parameters) and a strong use of data augmentation.}

\vspace{-3mm}
\subsection{WaveletUnet: wavelet-based models}
\vspace{-1mm}

WaveletUnet models rely on a cascade of two lazy, Haar or learnable wavelets to \mbox{downsample/upsample x4~(see section 2.4 and Fig.~4).} The ``{detail}" signals in Fig.~4 contain the skip connections' signal, and the ``{coarse}" signals are processed by the next layer.
In Table~1, we note that lazy wavelet outperforms Haar wavelet, and that learnable wavelet slighlty underperforms lazy wavelet. Further,~post-networks and normalization do not improve SDR results. Hence, we select the \mbox{base lazy WaveletUnet  for a listening test.} 
Finally, and similarly as for the WaveUnets above, the WaveletUnets we consider also obtatin much better results than the original one.

\vspace{-1.5mm}
\section{Subjective evaluation}
\vspace{-1mm}

\mbox{We further evaluate the most promising models with a listening test.}
10 expert listeners {evaluate 4 songs:} 2 from the Free Music Archive~\cite{defferrard2017fma}, and 2 from the test-set\footnote{\label{note1}Listen online: \url{http://jordipons.me/apps/upsamplers/}}. In this listening test, we evaluate ``overall separation quality"~\mbox{(from 1 to 5, the higher the better)} via averaging the ratings obtained from individually evaluating each estimated source (Table 2). 
WaveUnets obtain the best MOS ratings, and nearest neighbour is preferred over stretch (t-test: {\footnotesize $p$$<$$10^{-5}$}). This is possibly because 
%Informal listening\footref{note1} reveals that 
stretch can introduce tonal artifacts (due~to spectral replicas of offsets), but nearest neighbor cannot (because its filtering artifacts attenuate spectral replicas of offsets)\footref{note1}.~\mbox{This denotes} that filtering artifacts can be useful to combat the spectral replicas of signal offsets.
We also find that stretch is preferred over Demucs no overlap (t-test: {\footnotesize $p$$<$$10^{-8}$}), that Demucs no overlap is preferred over WaveShuffle + norm (t-test: {\footnotesize $p$$=$$0.03$}), and that WaveShuffle~+~norm and lazy WaveletUnet are \mbox{the worst rated (and perform equivalently,} t-test: {\footnotesize $p$$=$$0.79$}). 
%First, we  find that~(i) the WaveShuffle model, with best SDR scores, and the Lazy WaveletUnet 
%are the worst rated. 
Further, WaveShuffle~+~norm (with the best SDR scores) obtains among the worse MOS ratings, and WaveUnet nearest neighbor (with the best MOS ratings) obtains lower SDR scores. 
This shows that small SDR differences don't translate directly to perceptual improvement.  
Finally, we want to note that post-neworks and normalization \mbox{layers were unable to fully remove tonal artifacts.\footref{note1} 
}

\begin{table}
   \vspace{-1mm}
	\resizebox{\columnwidth}{!}{\begin{tabular}{| l || c c c c|}
			\hline	
			\textbf{Music Source Separation}  & {input} & {approx.} & SDR& {epoch}   \\
			\textit{{MUSDB's test-set results}}  & {(sec)} & {\# parm} &~(dB) & {(sec)}   \\
			\hline	\hline

			WaveUnet~\cite{stoller2018wave}: original publication & 6.7 & 10M & 3.23 & - \\
			
			WaveletUnet~\cite{nakamura2020time}: original publication & 6.7 & 15M & 3.39 & - \\
			
			Demucs~\cite{defossez2019music}: original publication & 10 & 648M & 5.34 & - \\

						\hline \hline
			
			Demucs: partial overlap & 12  & 716M & 5.28  &  316 \\
			
			Demucs: full overlap  & 12  & 703M & 5.37 &  311 \\ 
			
			{Demucs: no overlap} & {12} & {648M} &  \textbf{5.39} &  298 \\ 
			
			\hline
			
			Demucs: no overlap + post-networks & 12  & 648M  & \textbf{5.39}  & 296 \\ 
			
			{Demucs: no overlap + normalization }& 12  & 648M  & 5.30 & 299 \\ 
			
			\hline \hline

			{WaveShuffle}  & {12}  & {729M} & {5.38} & {305} \\
			
			\hline
			
			WaveShuffle + post-network & 12   & 729M & 5.38 & 302 \\
		
			{WaveShuffle + normalization }& 12  & 729M  &\textbf{ 5.44 } & 308 \\ 	
			
			\hline \hline

			{WaveUnet: sinc}  & 12  & 716M & 4.52 & 553 \\ 
			
			WaveUnet: linear  & 12 & 716M & 4.62 & {430} \\ 
			
			{WaveUnet: nearest neighbor~(NN)} & {12}  & {716M}  & \textbf{5.17} & {420} \\ 
			
			{WaveUnet: stretch}  & {12} & {716M} & \textbf{5.23}  & 423 \\ 
			
			\hline
			
			WaveUnet NN + post-network & 12  & 716M  & \textbf{5.17} & 422 \\			
			
			WaveUnet NN + normalization  & 12  & 716M  & 5.08  & 421\\ 		
					
			\hline
			
			WaveUnet stretch + post-network& 12  & 716M & 5.23 & 420 \\			
			
			{WaveUnet stretch + normalization }& {12}  & {716M } & \textbf{ 5.24 } & 429 \\ 				

			\hline\hline
			
			{WaveletUnet: lazy wavelet} & {12}  & {716M} & \textbf{5.31} & 532 \\
			% 532
			{WaveletUnet}: Haar wavelet & 12  & 716M & 4.55 & 534 \\

			{WaveletUnet}: learnable wavelet & 12 & 716M & 5.30 & 535 \\ 
			
			\hline
			Lazy WaveletUnet + post-network & {12}  & {716M} & \textbf{5.31} & 530 \\
			
			Lazy WaveletUnet + normalization & 12  & 716M  & 5.22  & 534 \\ 			\hline

	\end{tabular}}
	\vspace{-3mm}
	\caption{Benchmarking upsampling layers for source separation.}
	\vspace{-1mm}
	\label{tab:table-name}
\end{table}

\begin{table}
	\resizebox{\columnwidth}{!}{\begin{tabular}{|c||c|c|c|c|c|}
			\hline
		     &	{{Demucs:}} & {WaveShuffle} & {WaveUnet:} & {WaveUnet:} &  {WaveletUnet:}  \\
			 & {no overlap} & {+ norm}  & {nearest neigh.} & {stretch}  & {lazy wavelet} \\	 \hline \hline
			\textbf{MOS } $\uparrow$&	2.70 &  2.48 & \textbf{3.30} & 2.83 & 2.45 \\		
				\hline
			
	\end{tabular}}
	\vspace{-2.5mm}	
	\caption{Mean opinion scores~(MOS, from 1 to 5) reporting overall separation quality of the most promising upsampling layers.}
	\vspace{-3mm}	
\end{table}

\vspace{-2mm}
\section{Conclusions}
\vspace{-1mm}

{Tonal artifacts are unpleasant artifacts difficult to tame because two causing factors are simultaneously interacting: the architecture~(as for transposed and subpixel convolutions) and the spectral replicas. 
Further, filtering artifacts are perceptually managable and a convenient tool against tonal artifacts. %---provided that post-neworks and normalization layers were unable to fully remove tonal artifacts.} 
We note this in our listening test, where nearest neighbor layers were preferred, possibly because these are free of tonal artifacts~(since it attenuates bands with spectral replicas of offsets) and allow high-frequency replicas~(yielding coherent high-frequency content for wide-band synthesis). 
%(what is beneficial, but challenging for sinc or linear). 
Yet, our results also show that many upsamplers can perform comparably (MOS$\approx$3) and, depending on our goals~(e.g., low memory footprint or no tonal artifacts), it should be possible to replace one by another.}

\bibliographystyle{IEEEbib}
\bibliography{mybib}

\begin{thebibliography}{10}

\bibitem{donahue2018adversarial}
Chris Donahue, Julian McAuley, and Miller Puckette,
\newblock ``Adversarial audio synthesis,''
\newblock in {\em ICLR}, 2019.

\bibitem{stoller2018wave}
Daniel Stoller, Sebastian Ewert, and Simon Dixon,
\newblock ``Wave-u-net: A multi-scale neural network for end-to-end audio
  source separation,''
\newblock in {\em ISMIR}, 2018.

\bibitem{pons2020upsampling}
Jordi Pons, Santiago Pascual, Giulio Cengarle, and Joan Serr{\`a},
\newblock ``Upsampling artifacts in neural audio synthesis,''
\newblock in {\em ICASSP}, 2020.

\bibitem{pandey2020densely}
Ashutosh Pandey and DeLiang Wang,
\newblock ``Densely connected neural network with dilated convolutions for
  real-time speech enhancement in the time domain,''
\newblock in {\em ICASSP}, 2020.

\bibitem{kumar2019melgan}
Kundan Kumar, Rithesh Kumar, Thibault de~Boissiere, Lucas Gestin, Wei~Zhen
  Teoh, Jose Sotelo, Alexandre de~Br{\'e}bisson, Yoshua Bengio, and Aaron~C
  Courville,
\newblock ``Melgan: Generative adversarial networks for conditional waveform
  synthesis,''
\newblock in {\em NeurIPS}, 2019.

\bibitem{defossez2019music}
Alexandre D{\'e}fossez, Nicolas Usunier, L{\'e}on Bottou, and Francis Bach,
\newblock ``Music source separation in the waveform domain,''
\newblock in {\em arXiv}, 2019.

\bibitem{gritsenko2020spectral}
Alexey~A Gritsenko, Tim Salimans, Rianne van~den Berg, Jasper Snoek, and Nal
  Kalchbrenner,
\newblock ``A spectral energy distance for parallel speech synthesis,''
\newblock in {\em arXiv}, 2020.

\bibitem{nakamura2020time}
Tomohiko Nakamura and Hiroshi Saruwatari,
\newblock ``Time-domain audio source separation based on wave-u-net combined
  with discrete wavelet transform,''
\newblock in {\em ICASSP}, 2020.

\bibitem{dhariwal2020jukebox}
Prafulla Dhariwal, Heewoo Jun, Christine Payne, Jong~Wook Kim, Alec Radford,
  and Ilya Sutskever,
\newblock ``Jukebox: A generative model for music,''
\newblock in {\em arXiv}, 2020.

\bibitem{odena2016deconvolution}
Augustus Odena, Vincent Dumoulin, and Chris Olah,
\newblock ``Deconvolution and checkerboard artifacts,''
\newblock {\em Distill}, vol. 1, no. 10, pp. e3, 2016.

\bibitem{pascual2017segan}
Santiago Pascual, Antonio Bonafonte, and Joan Serr{\`a},
\newblock ``Segan: Speech enhancement generative adversarial network,''
\newblock in {\em Interspeech}, 2017.

\bibitem{germain2019speech}
Francois~G. Germain, Qifeng Chen, and Vladlen Koltun,
\newblock ``Speech denoising with deep feature losses,''
\newblock in {\em Interspeech}, 2019.

\bibitem{smith2007mathematics}
Julius~O. Smith,
\newblock ``Mathematics of the discrete fourier transform (dft): with audio
  applications,'' 2007.

\bibitem{shi2016real}
Wenzhe Shi, Jose Caballero, Ferenc Husz{\'a}r, Johannes Totz, Andrew~P Aitken,
  Rob Bishop, Daniel Rueckert, and Zehan Wang,
\newblock ``Real-time single image and video super-resolution using an
  efficient sub-pixel convolutional neural network,''
\newblock in {\em CVPR}, 2016.

\bibitem{aitken2017checkerboard}
Andrew Aitken, Christian Ledig, Lucas Theis, Jose Caballero, Zehan Wang, and
  Wenzhe Shi,
\newblock ``Checkerboard artifact free sub-pixel convolution: A note on
  sub-pixel convolution, resize convolution and convolution resize,''
\newblock in {\em arXiv}, 2017.

\bibitem{sweldens1998lifting}
Wim Sweldens,
\newblock ``The lifting scheme: A construction of second generation wavelets,''
\newblock {\em {SIAM Journal on Mathematical Analysis}}, vol. 29, no. 2, pp.
  511--546, 1998.

\bibitem{musdb18}
Zafar Rafii, Antoine Liutkus, Fabian-Robert St{\"o}ter, Stylianos~Ioannis
  Mimilakis, and Rachel Bittner,
\newblock ``The {MUSDB18} corpus for music separation,'' 2017.

\bibitem{vincent2006performance}
Emmanuel Vincent, R{\'e}mi Gribonval, and C{\'e}dric F{\'e}votte,
\newblock ``Performance measurement in blind audio source separation,''
\newblock {\em IEEE TASLP}, vol. 14, no. 4, pp. 1462--1469, 2006.

\bibitem{dauphin2017language}
Yann~N Dauphin, Angela Fan, Michael Auli, and David Grangier,
\newblock ``Language modeling with gated convolutional networks,''
\newblock in {\em ICML}, 2017.

\bibitem{uhlich2017improving}
Stefan Uhlich, Marcello Porcu, Franck Giron, Michael Enenkl, Thomas Kemp, Naoya
  Takahashi, and Yuki Mitsufuji,
\newblock ``Improving music source separation based on deep neural networks
  through data augmentation and network blending,''
\newblock in {\em ICASSP}, 2017.

\bibitem{ulyanov2016instance}
Dmitry Ulyanov, Andrea Vedaldi, and Victor Lempitsky,
\newblock ``Instance normalization: The missing ingredient for fast
  stylization,''
\newblock in {\em arXiv}, 2016.

\bibitem{kingma2014adam}
Diederik~P Kingma and Jimmy Ba,
\newblock ``Adam: A method for stochastic optimization,''
\newblock in {\em arXiv}, 2014.

\bibitem{defferrard2017fma}
Micha{\"e}l Defferrard, Kirell Benzi, Pierre Vandergheynst, and Xavier Bresson,
\newblock ``{FMA: A Dataset For Music Analysis},''
\newblock in {\em ISMIR}, 2017.

\end{thebibliography}

\appendix
\section{Stretch interpolation:\\frequency response derivation }
{Stretch interpolation} upsamples the signal with zeros as follows:
\[x_{\text{stretch by} M}[n]=\sum_{k=-\infty}^{\infty} x[k] \delta[n-kM],\]
\noindent where $M$ is the upsampling factor. Hence, its frequency response is:

\[X_{\text{stretch by} M}({\rm e}^{j\omega}) = X({\rm e}^{j\omega M}).\]

% https://www.youtube.com/watch?v=kKb_9rmTnHY&feature=emb_title

\section{Additional figures}

Fig. 8 is included because it helps understanding how spectral replicas and filtering artifacts interact, and we depict it alongside Figs. 6 and 7 (the exact same ones as Figs. 1 and 2) to facilitate comparing among those. Note that these three figures together are an interesting resource to understand how filtering artifacts, tonal artifacts and spectral replicas interact. {We recommend visualizing those while reading section 2.5.}

\begin{figure*}[t]
	\vfill \vspace{-9mm}
	\hspace{18mm}
	\begin{minipage}[b]{0.18\linewidth}
		\centering
		\centerline{\includegraphics[width=1.1\linewidth]{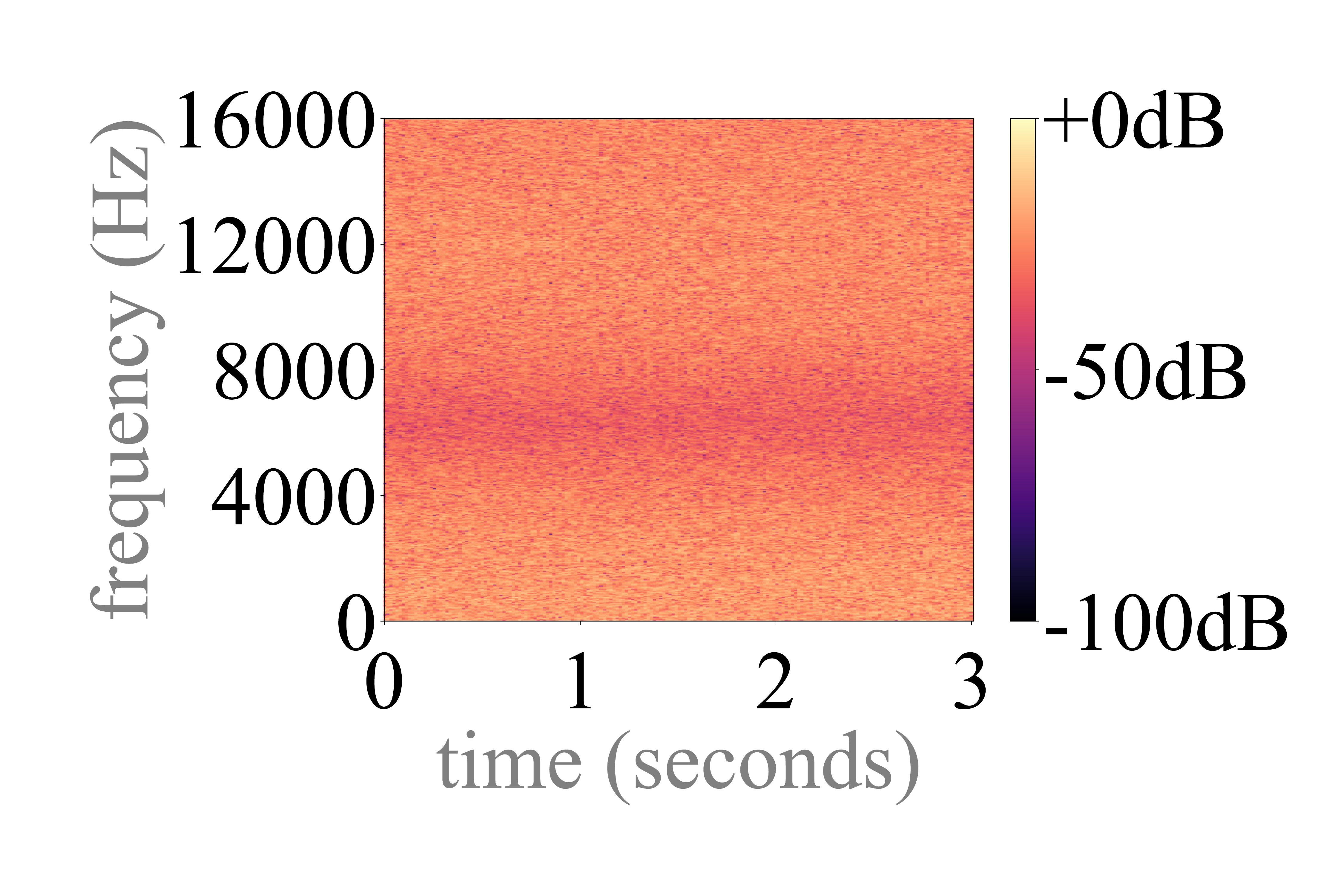}}
		\vspace{-3mm}
		{(a) Transposed CNN:\\no overlap (length=8)}\medskip
	\end{minipage}	
	\begin{minipage}[b]{0.18\linewidth}
		\centering
		\centerline{\includegraphics[width=1.1\linewidth]{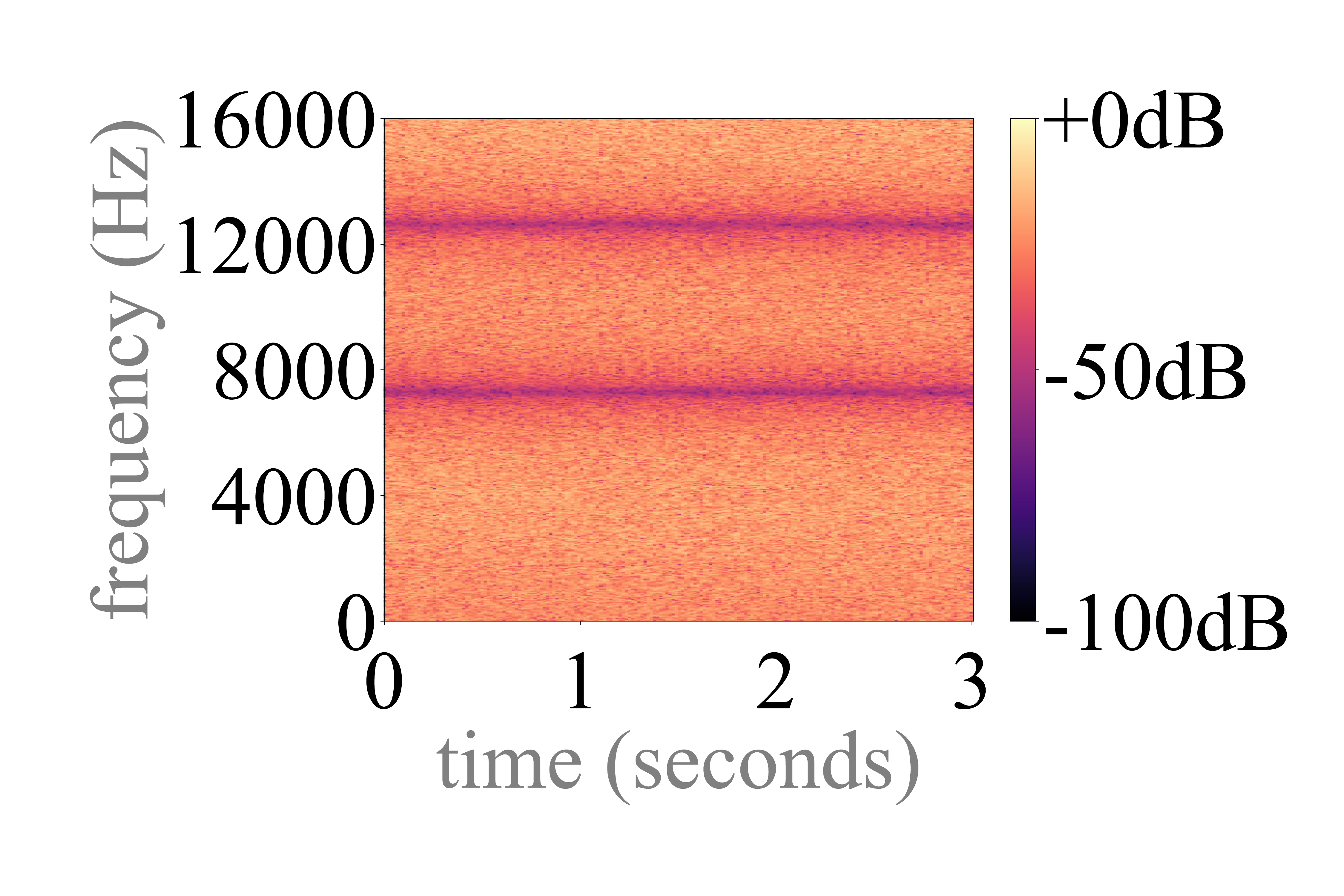}}
		\vspace{-3mm}
		{(b) Transposed CNN:\\\mbox{partial overlap (length=9)}}\medskip
	\end{minipage}
	\begin{minipage}[b]{0.18\linewidth}
		\centering
		\centerline{\includegraphics[width=1.1\linewidth]{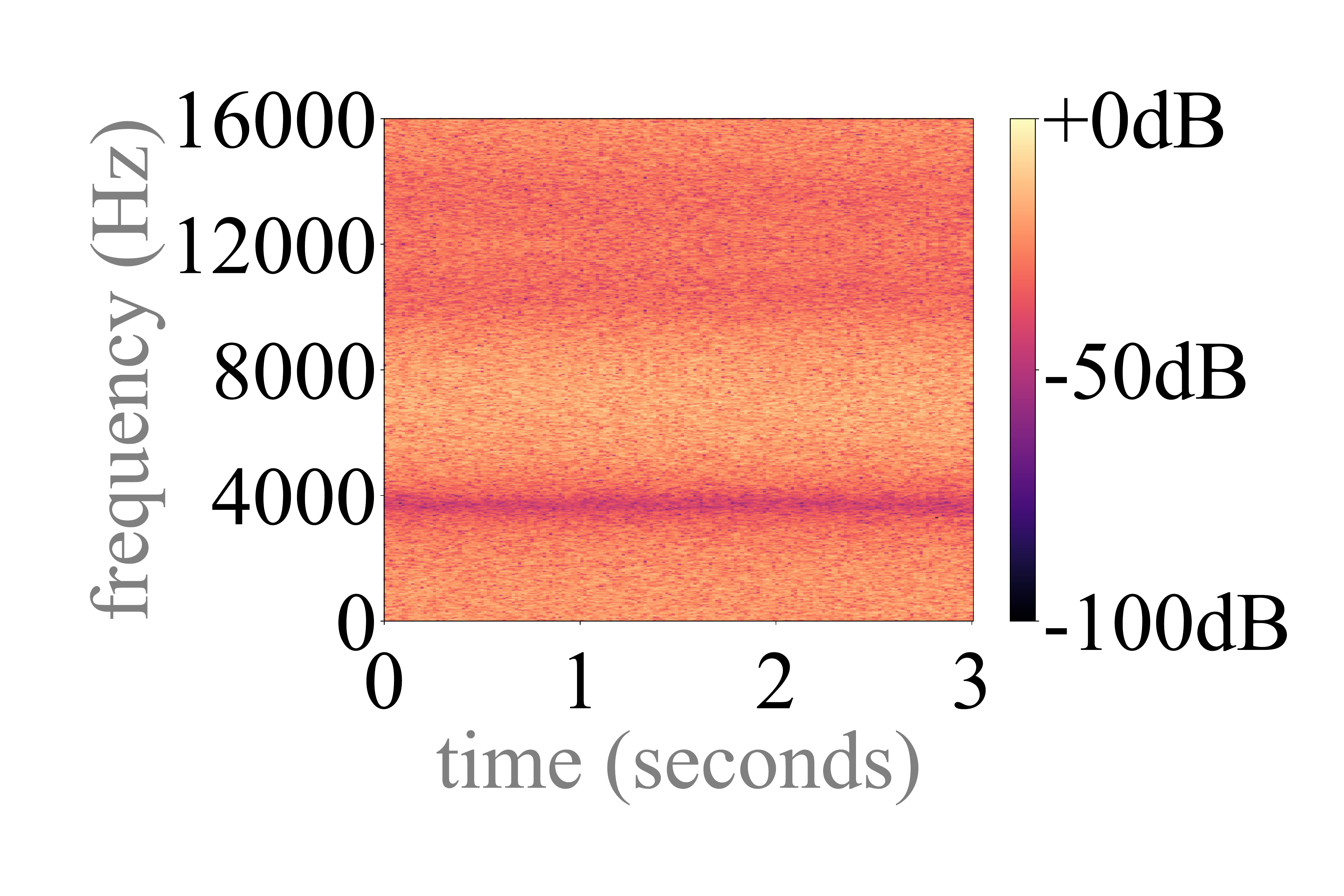}}
		\vspace{-3mm}
		{(c) Transposed CNN:\\full overlap (length=4)}\medskip
	\end{minipage}
	\begin{minipage}[b]{0.18\linewidth}
		\centering
		\centerline{\includegraphics[width=1.1\linewidth]{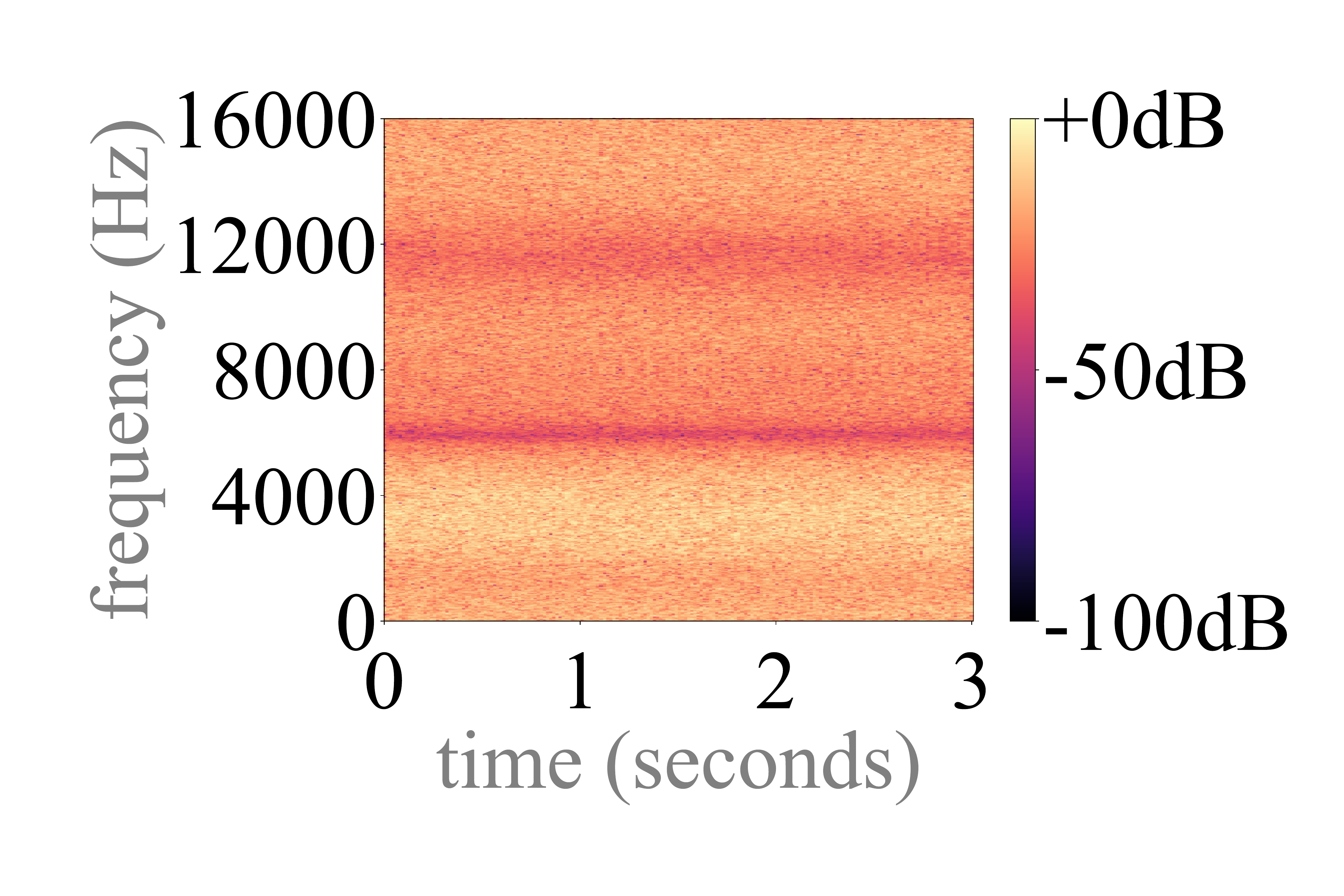}}
		\vspace{-3mm}
		{(d) Subpixel CNN\\(filter length=9)}\medskip
	\end{minipage}
	\vfill \vspace{-4mm}
	\hspace{18mm}
	\begin{minipage}[b]{0.18\linewidth}
		\centering
		\centerline{\includegraphics[width=1.1\linewidth]{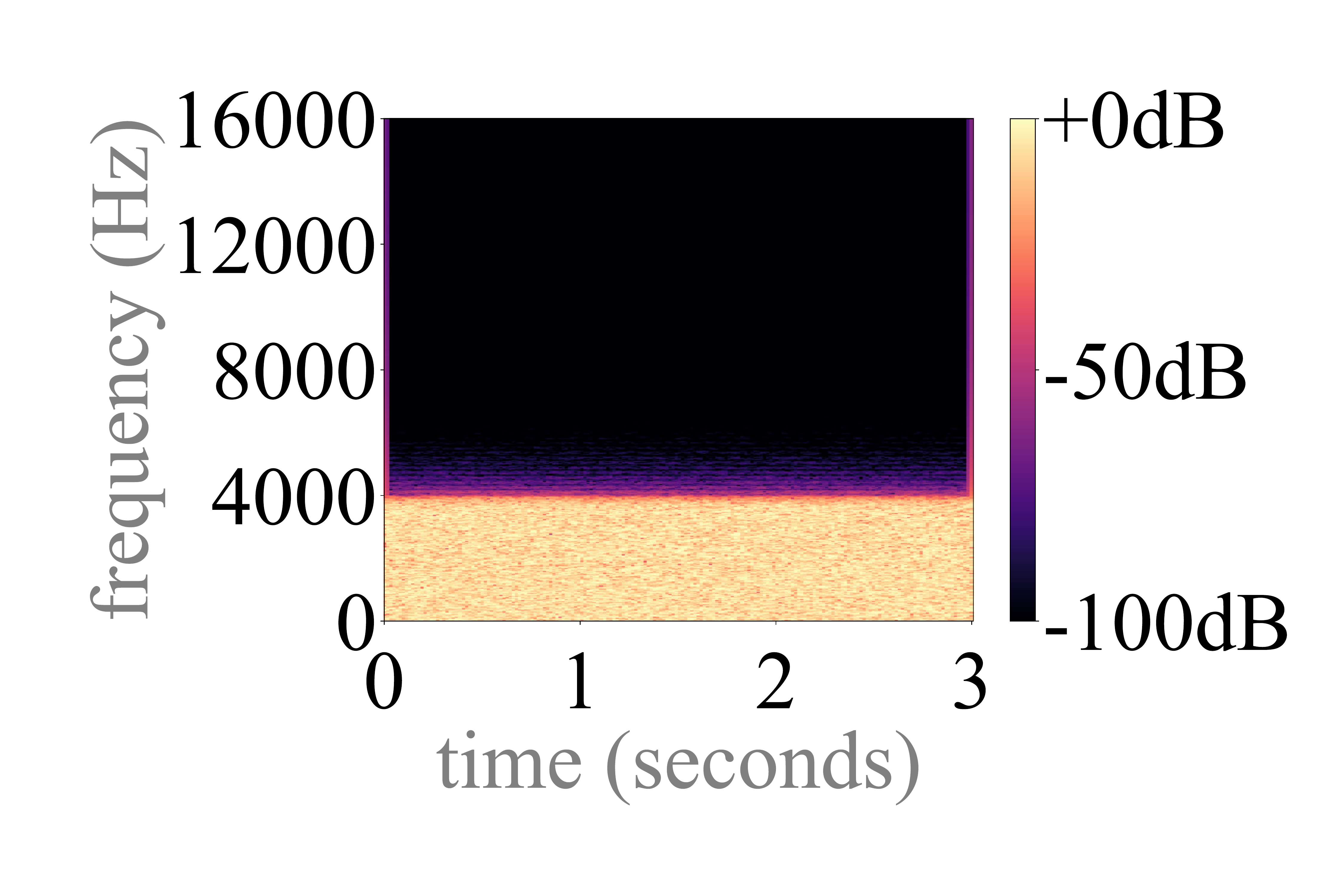}}
		\vspace{-3mm}
		{(e) Interpolation:\\sinc}\medskip
	\end{minipage}	
	\begin{minipage}[b]{0.18\linewidth}
		\centering
		\centerline{\includegraphics[width=1.1\linewidth]{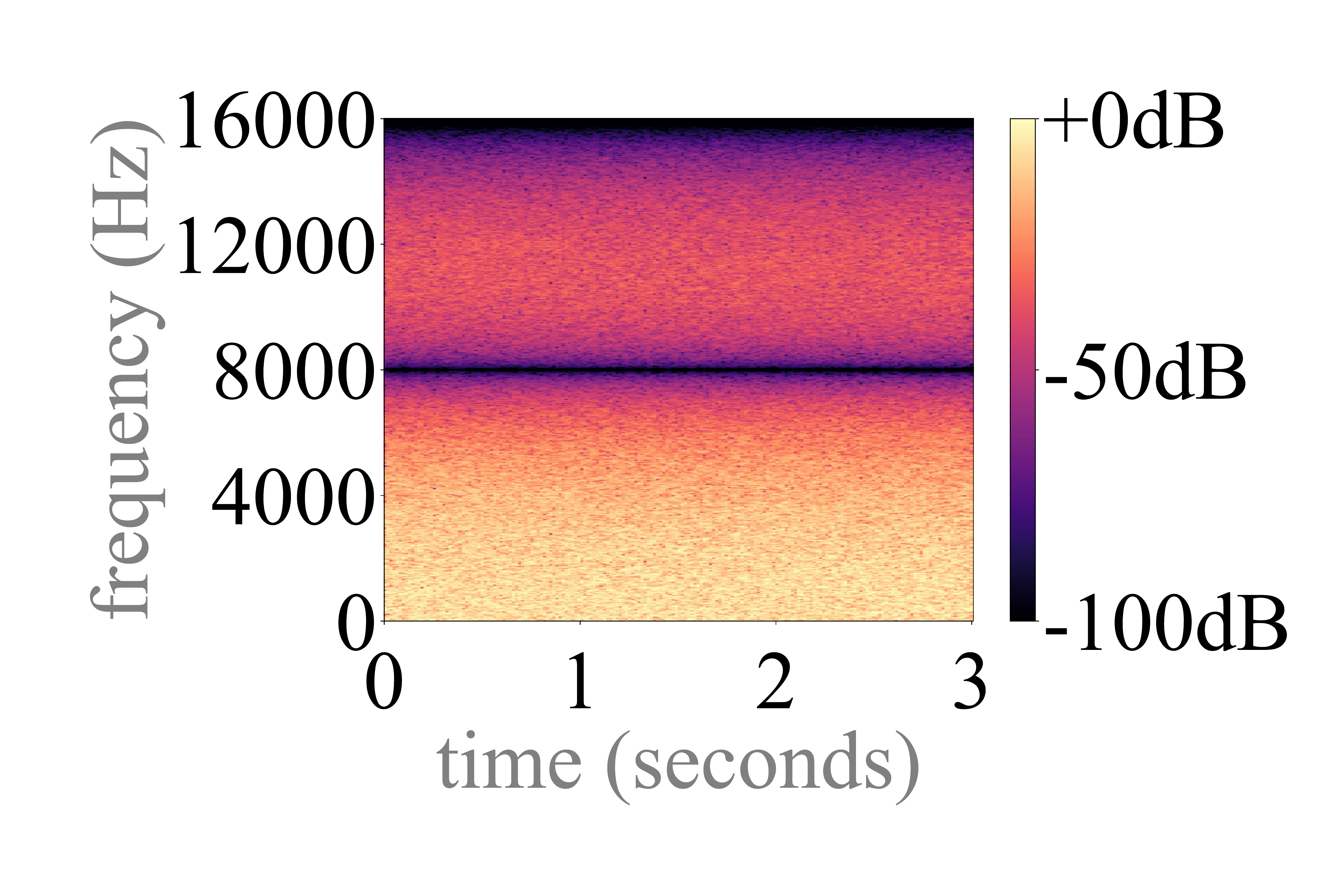}}
		\vspace{-3mm}
		{(f) Interpolation:\\linear}\medskip
	\end{minipage}
	\begin{minipage}[b]{0.18\linewidth}
		\centering
		\centerline{\includegraphics[width=1.1\linewidth]{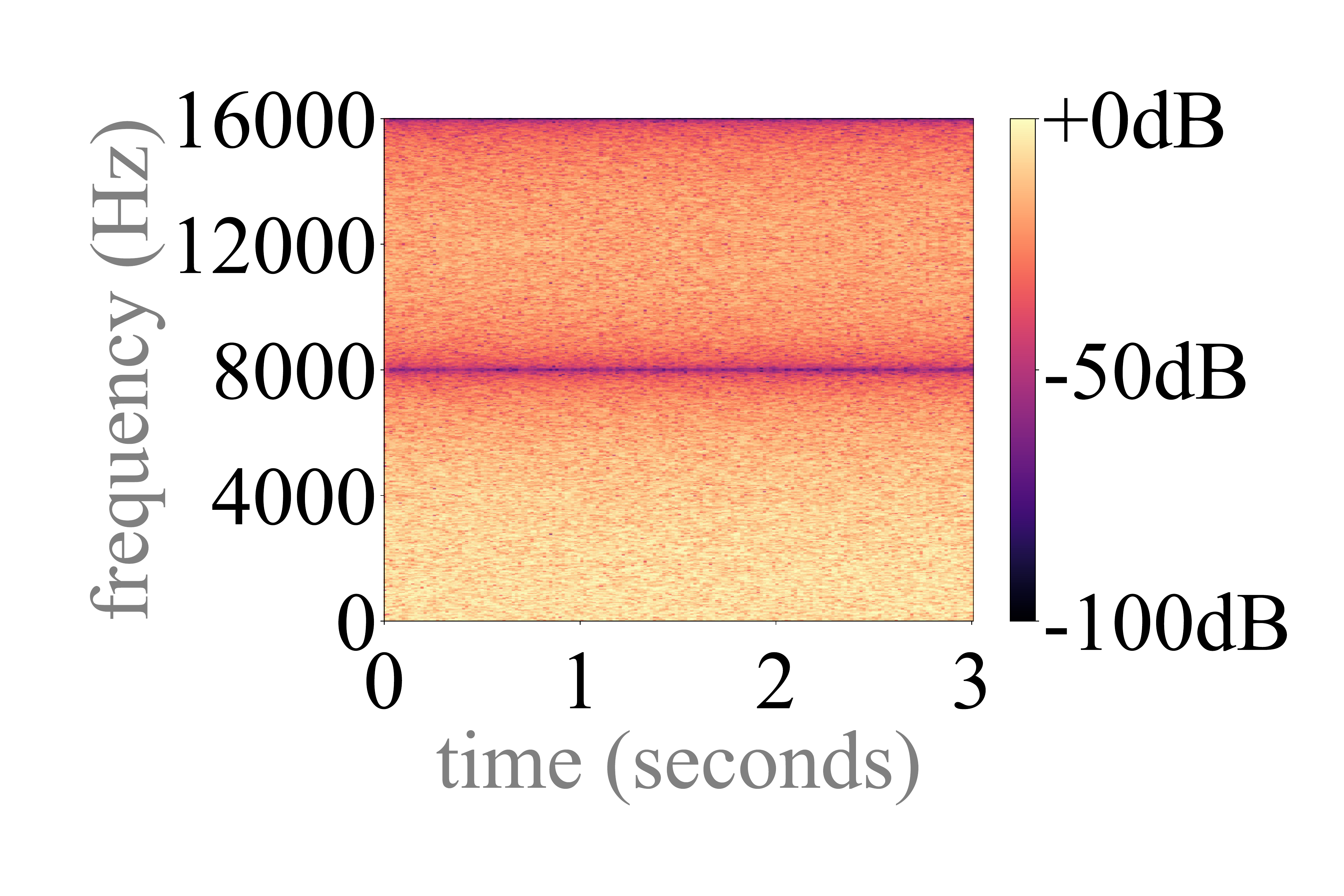}}
		\vspace{-3mm}
		{(g) Interpolation:\\nearest neighbor}\medskip
	\end{minipage}
	\begin{minipage}[b]{0.18\linewidth}
		\centering
		\centerline{\includegraphics[width=1.1\linewidth]{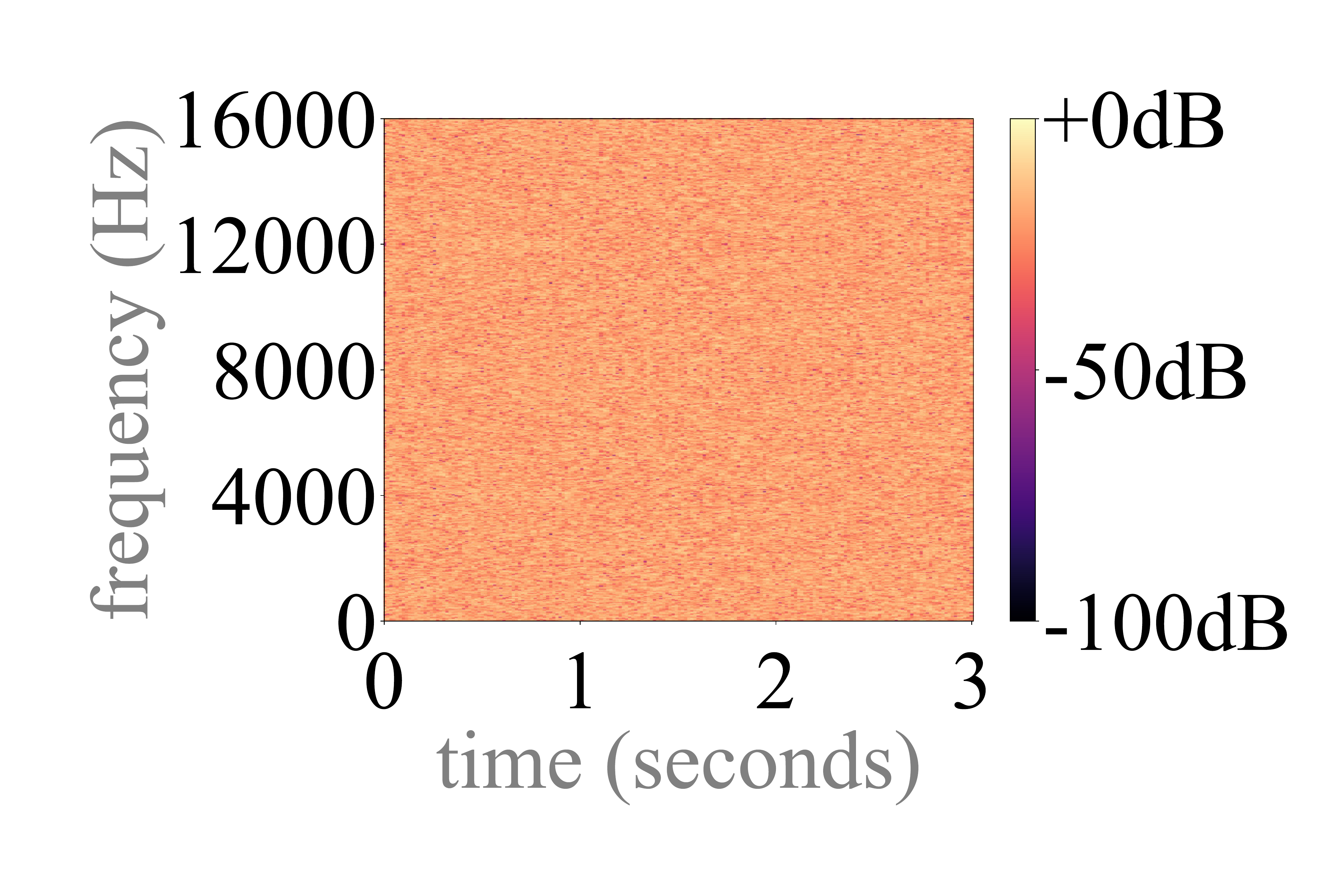}}
		\vspace{-3mm}
		{(h) Interpolation:\\stretch}\medskip
	\end{minipage}
	\vfill \vspace{-5mm}
	\caption{{Input: white noise at 8kHz. Upsampling~($\uparrow$4) layers  can introduce filtering artifacts that attenuate some bands. Only~(e, f, g) ``horizontal valleys" are considered filtering artifacts, because are caused by non-learnable interpolations. Hence,~(e, f, g) layers would introduce filtering  artifacts even after training---while the rest of ``horizontal valleys"~(a, b, c, d) can change during training. Transposed CNNs with stride=4. This figure is exactly the same as Fig. 1.}}
	\vspace{-4mm}
	\label{fig:all}
\end{figure*}

\begin{figure*}[t]
	\hspace{18mm}
	\begin{minipage}[b]{0.18\linewidth}
		\centering
		\centerline{\includegraphics[width=1.1\linewidth]{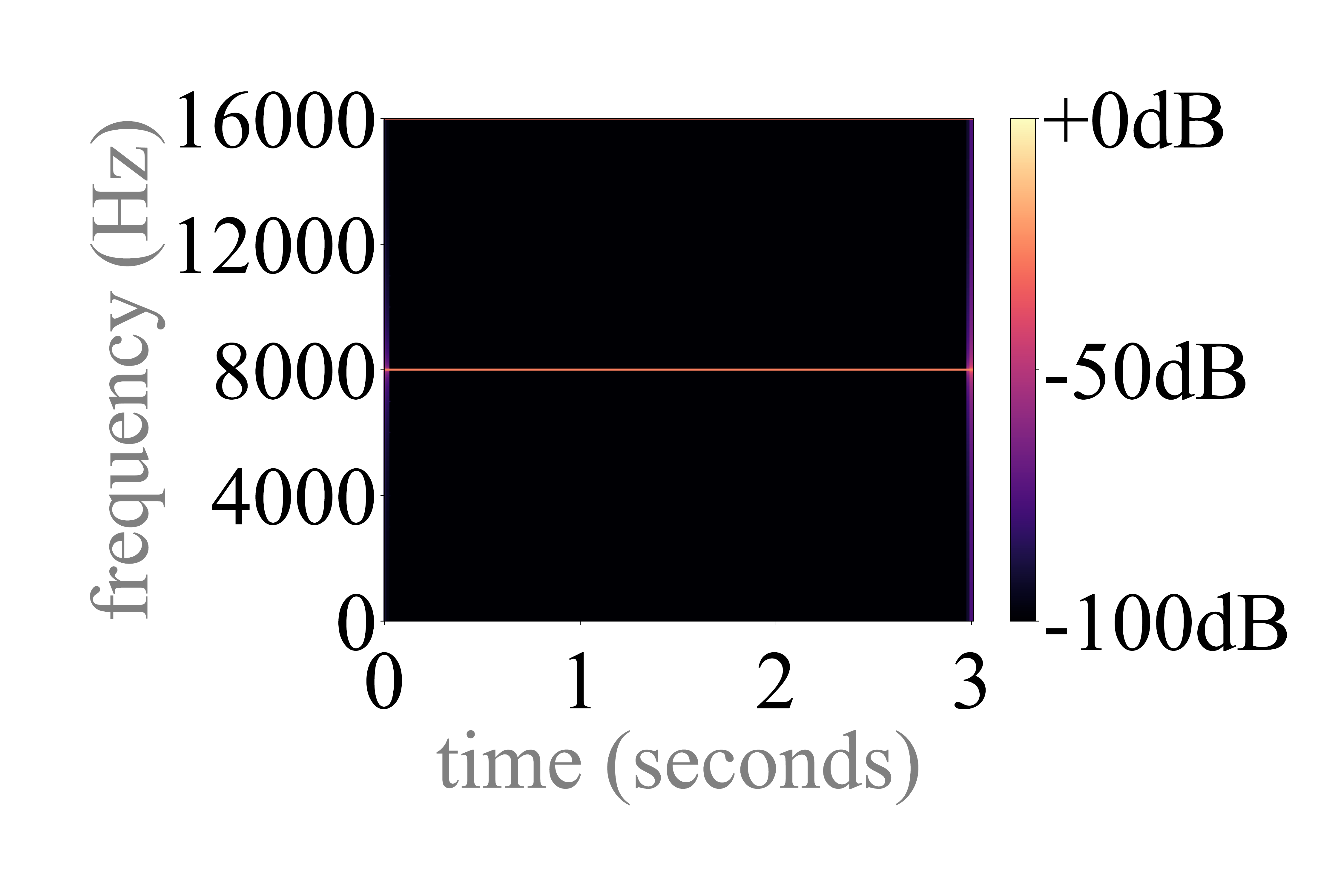}}
		\vspace{-3mm}
		{(a) Transposed CNN:\\no overlap (length=8)}\medskip
	\end{minipage}	
	\begin{minipage}[b]{0.18\linewidth}
		\centering
		\centerline{\includegraphics[width=1.1\linewidth]{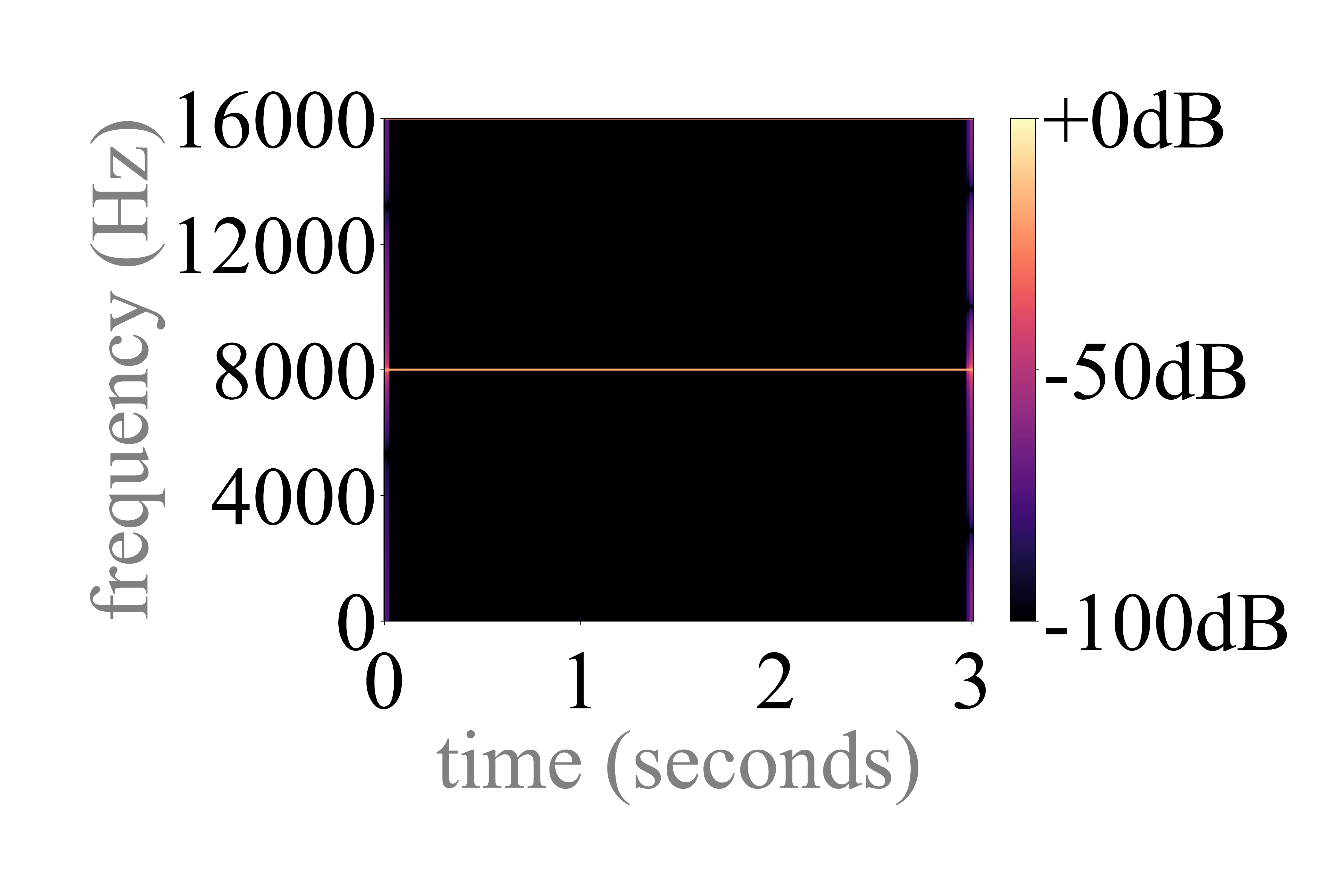}}		\vspace{-3mm}
		{(b) Transposed CNN:\\\mbox{partial overlap (length=9)}}\medskip
	\end{minipage}
	\begin{minipage}[b]{0.18\linewidth}
		\centering
		\centerline{\includegraphics[width=1.1\linewidth]{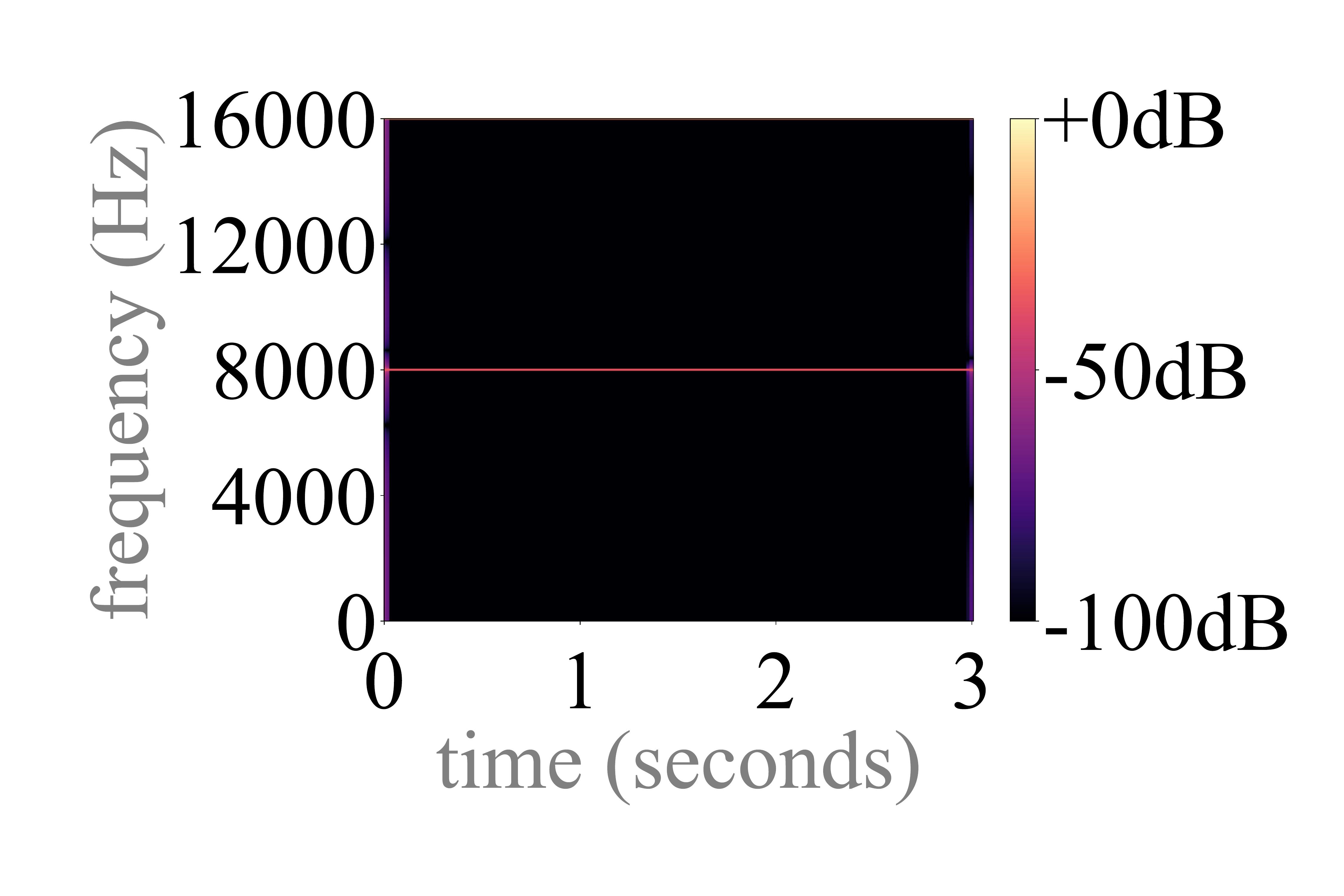}}		\vspace{-3mm}
		{(c) Transposed CNN:\\full overlap (length=4)}\medskip
	\end{minipage}
	\begin{minipage}[b]{0.18\linewidth}
		\centering
		\centerline{\includegraphics[width=1.1\linewidth]{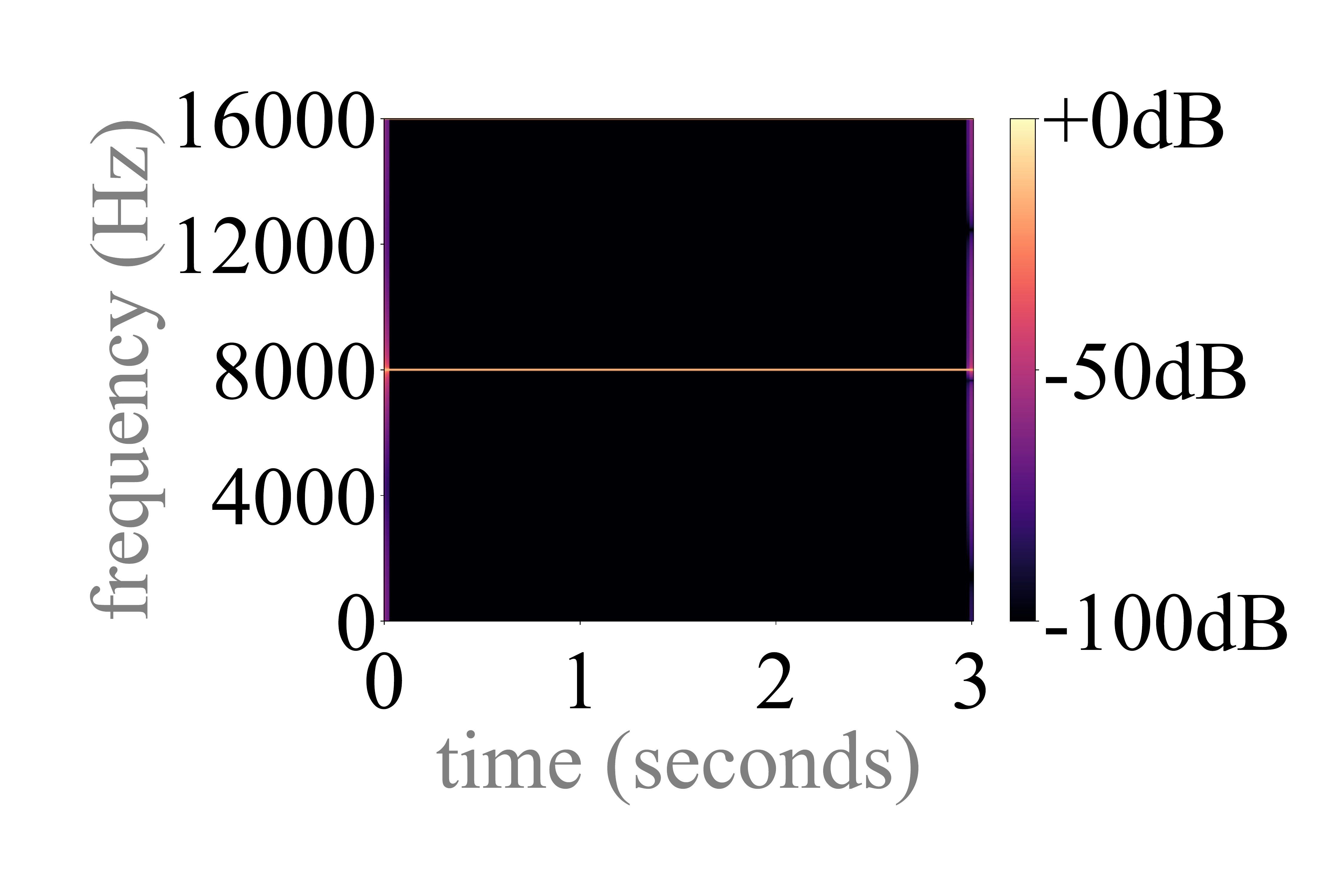}}		\vspace{-3mm}
		{(d) Subpixel CNN\\(filter length=9)}\medskip
	\end{minipage}
	\vfill \vspace{-4mm}
	\hspace{18mm}
	\begin{minipage}[b]{0.18\linewidth}
		\centering
		\centerline{\includegraphics[width=1.1\linewidth]{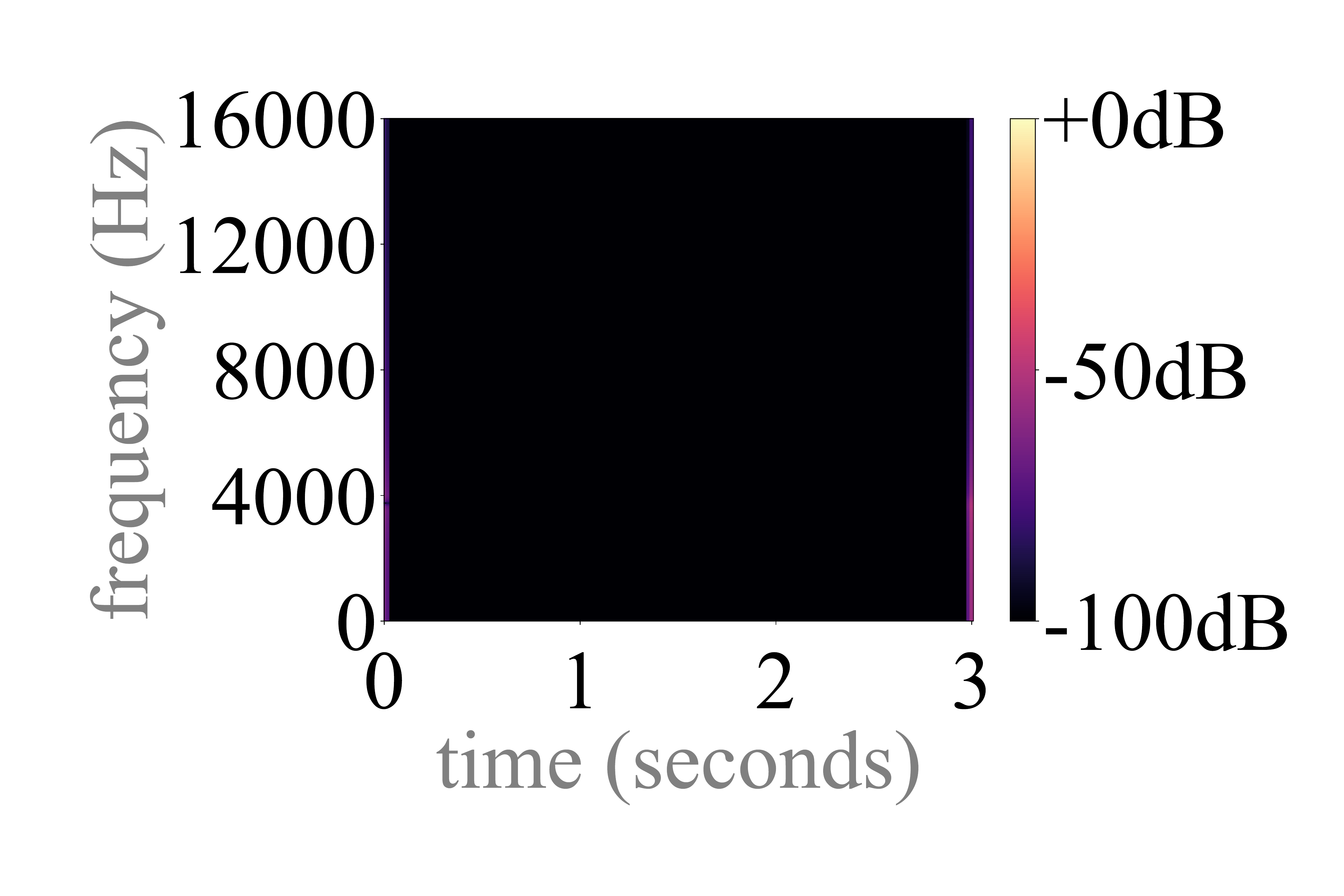}}		\vspace{-3mm}
		{(e) Interpolation:\\sinc}\medskip
	\end{minipage}	
	\begin{minipage}[b]{0.18\linewidth}
		\centering
		\centerline{\includegraphics[width=1.1\linewidth]{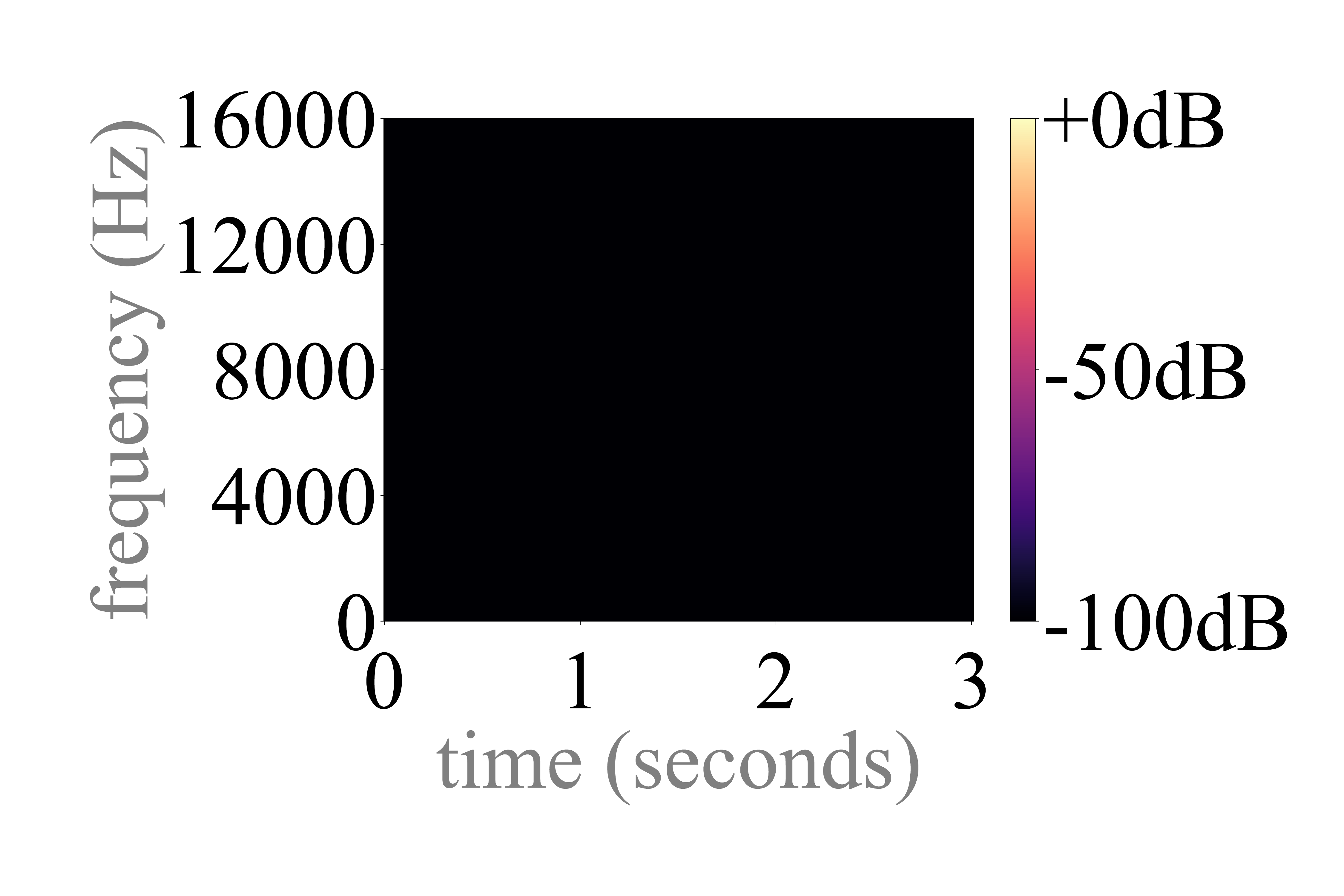}}		\vspace{-3mm}
		{(f) Interpolation:\\linear}\medskip
	\end{minipage}
	\begin{minipage}[b]{0.18\linewidth}
		\centering
		\centerline{\includegraphics[width=1.1\linewidth]{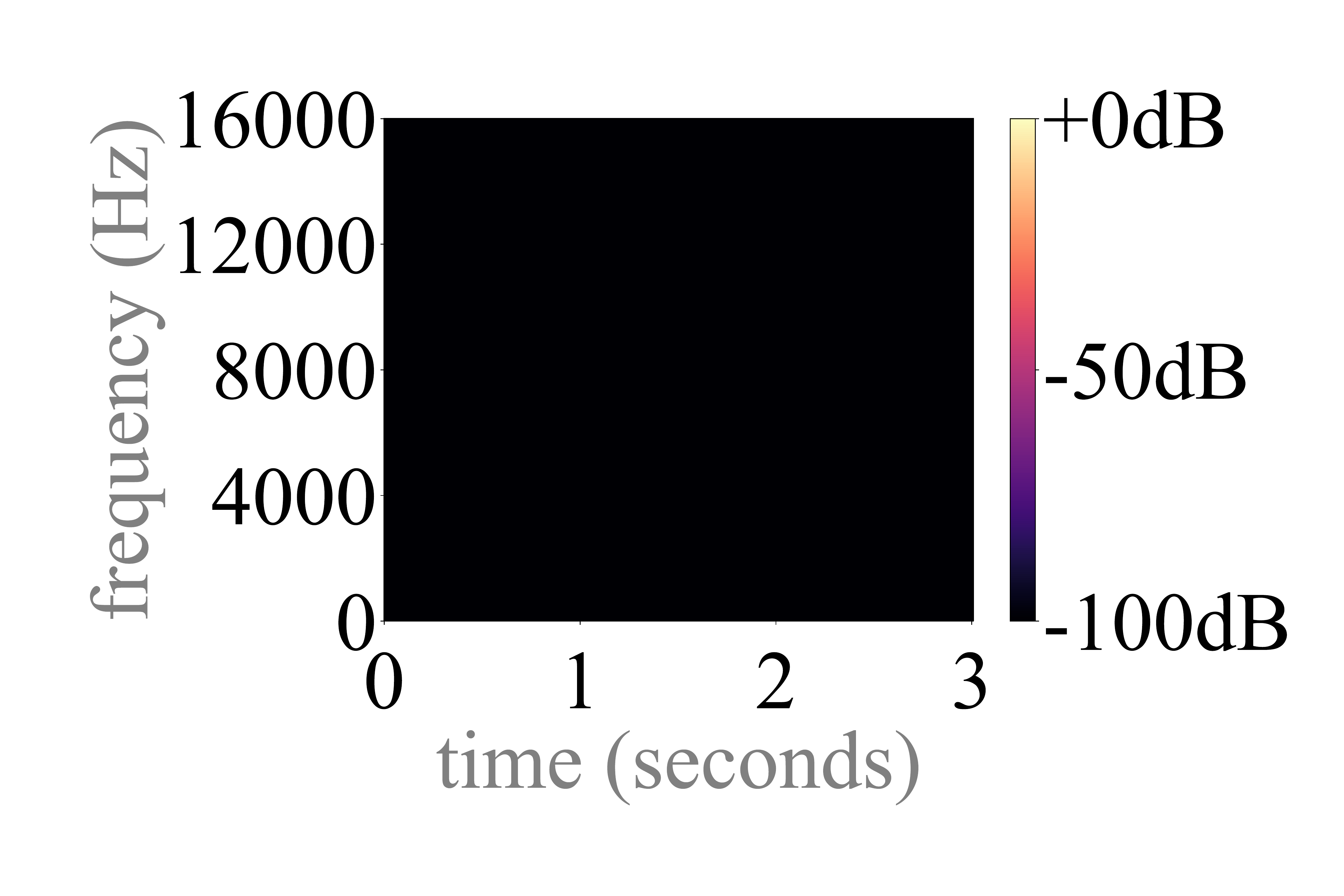}}		\vspace{-3mm}
		{(g) Interpolation:\\nearest neighbor}\medskip
	\end{minipage}
	\begin{minipage}[b]{0.18\linewidth}
		\centering
		\centerline{\includegraphics[width=1.1\linewidth]{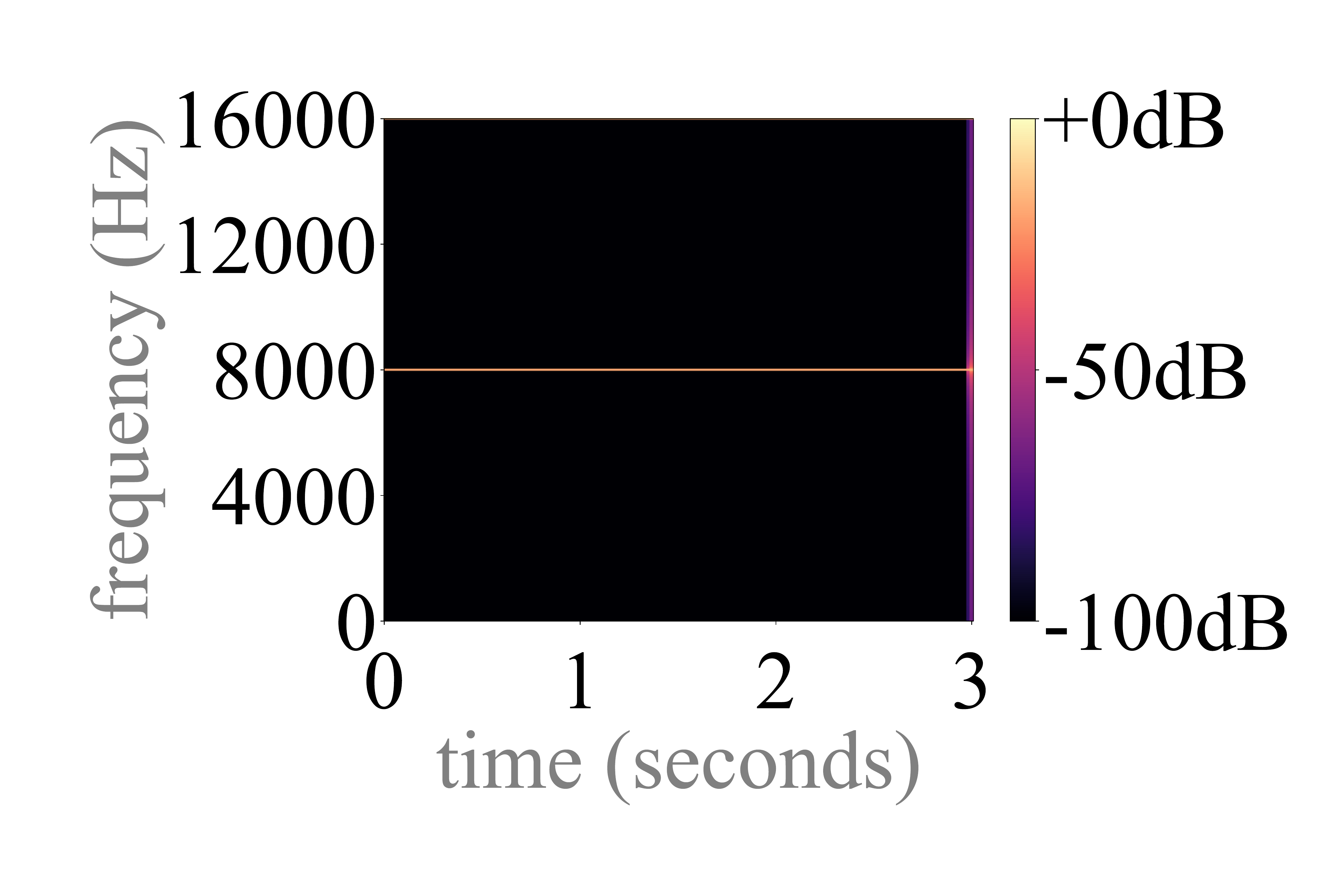}}		\vspace{-3mm}
		{(h) Interpolation:\\stretch}\medskip
	\end{minipage}
	\vfill \vspace{-5mm}
	\caption{{Input: ones~(constant) at 8kHz. Upsampling~($\uparrow$4) layers can introduce tonal artifacts~(horizontal lines).  Transposed CNNs with stride=4. This figure is exactly the same as Fig. 2.}}
	\vspace{-6mm}
	\label{fig:all}
\end{figure*}

\begin{figure*}[t]
	\hspace{18mm}
	\begin{minipage}[b]{0.18\linewidth}
		\centering
		\centerline{\includegraphics[width=1.1\linewidth]{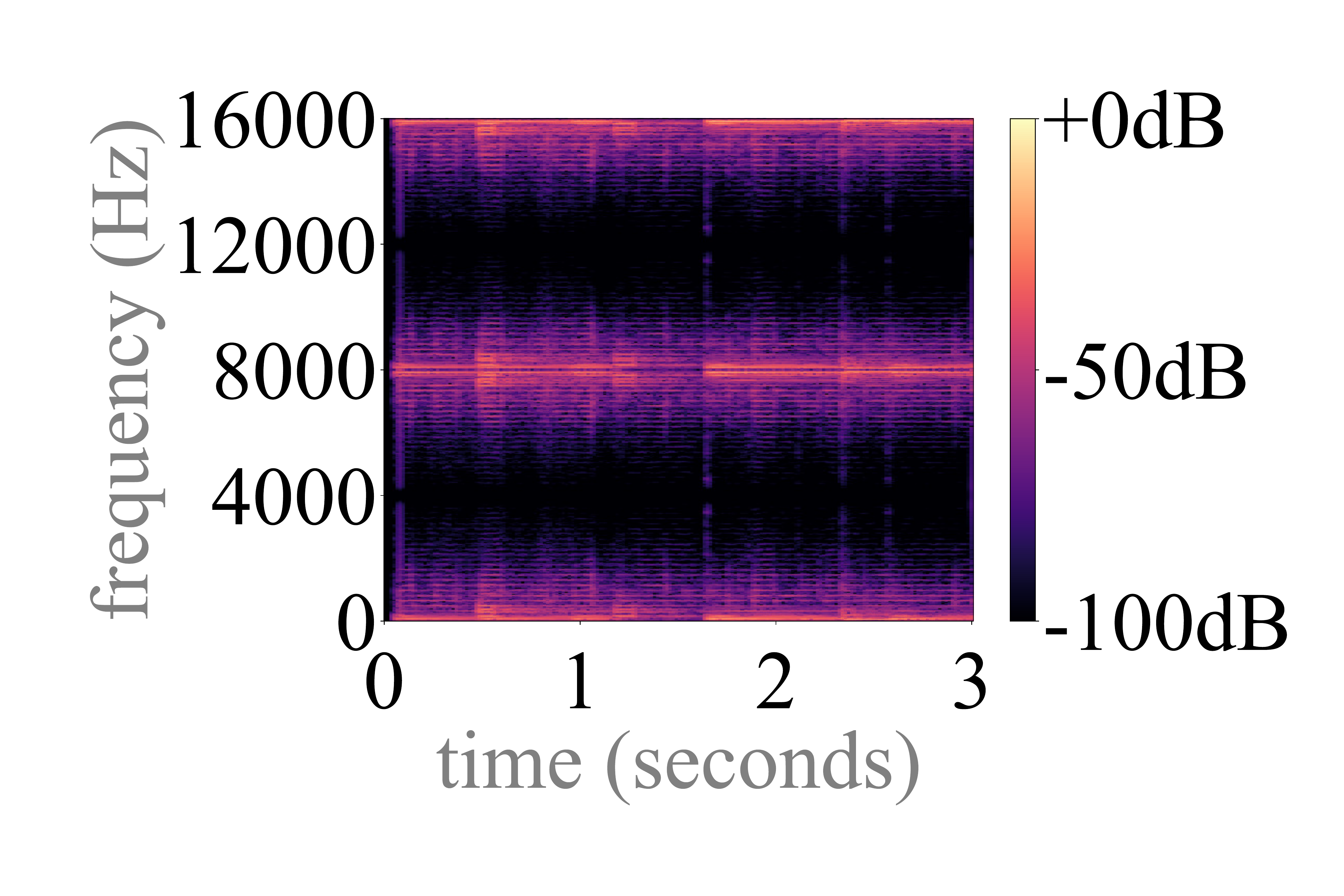}}
		\vspace{-3mm}
		{(a) Transposed CNN:\\no overlap (length=8)}\medskip
	\end{minipage}	
	\begin{minipage}[b]{0.18\linewidth}
		\centering
		\centerline{\includegraphics[width=1.1\linewidth]{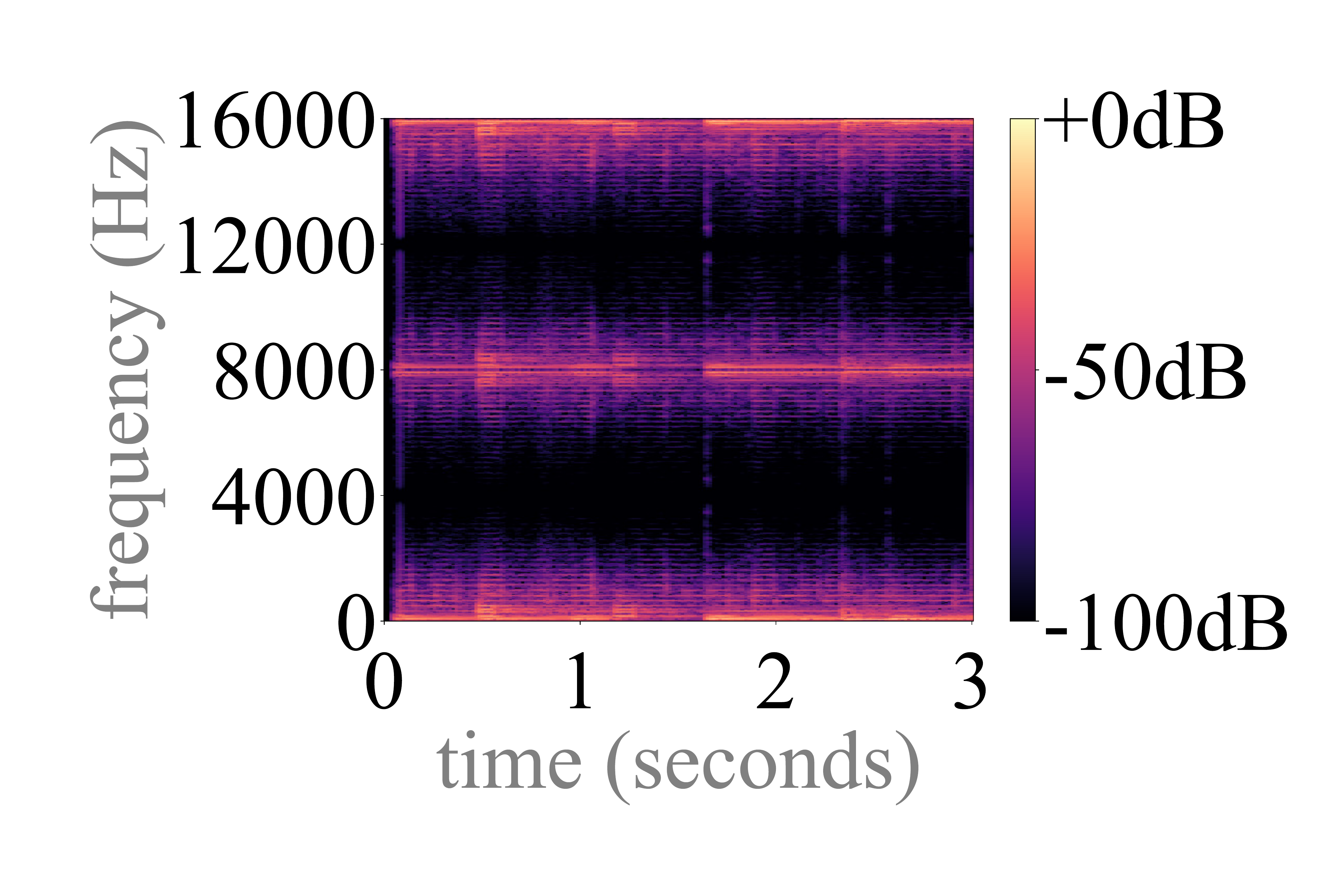}}		\vspace{-2mm}
		{(b) Transposed CNN:\\\mbox{partial overlap (length=9)}}\medskip
	\end{minipage}
	\begin{minipage}[b]{0.18\linewidth}
		\centering
		\centerline{\includegraphics[width=1.1\linewidth]{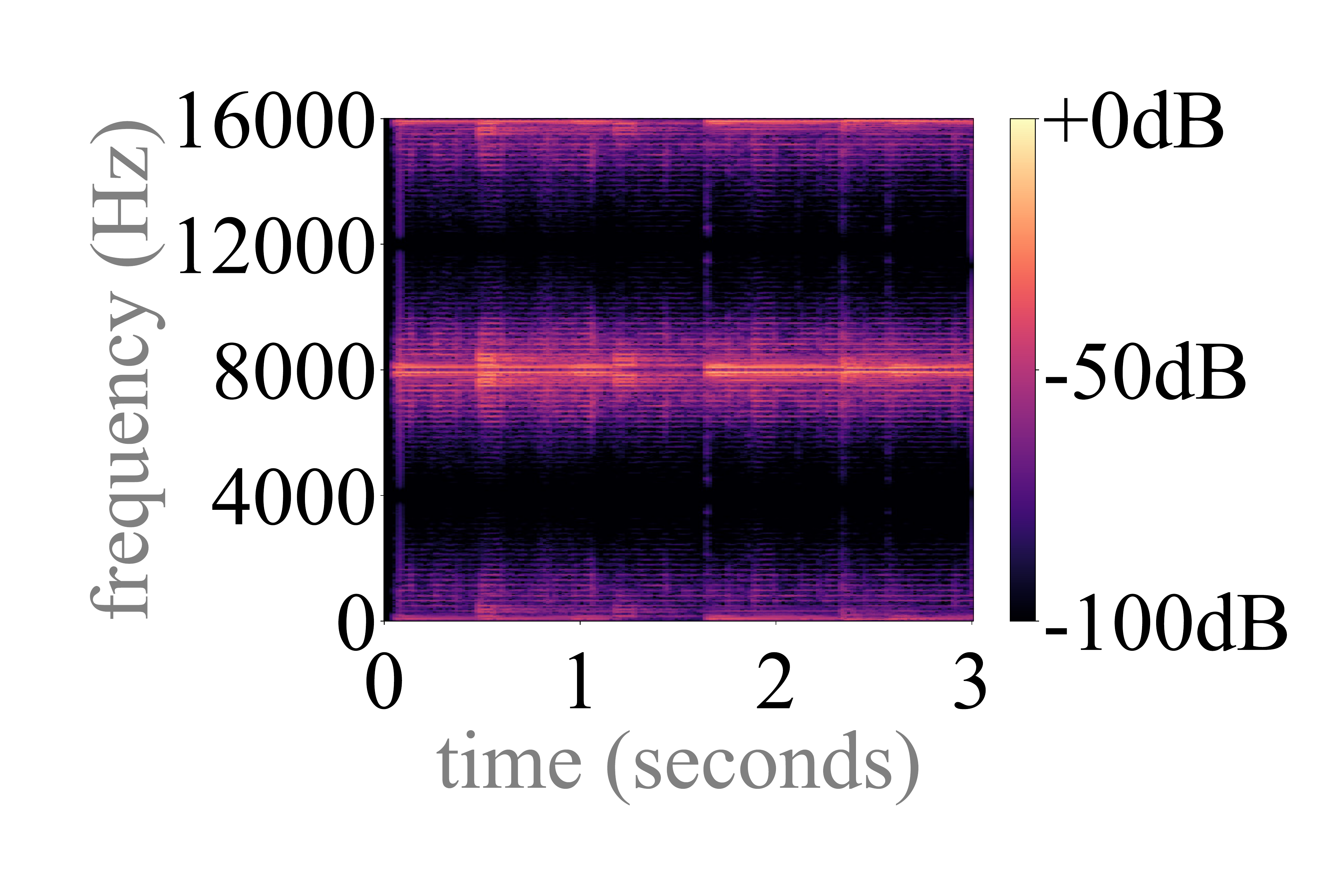}}		\vspace{-2mm}
		{(c) Transposed CNN:\\full overlap (length=4)}\medskip
	\end{minipage}
	\begin{minipage}[b]{0.18\linewidth}
		\centering
		\centerline{\includegraphics[width=1.1\linewidth]{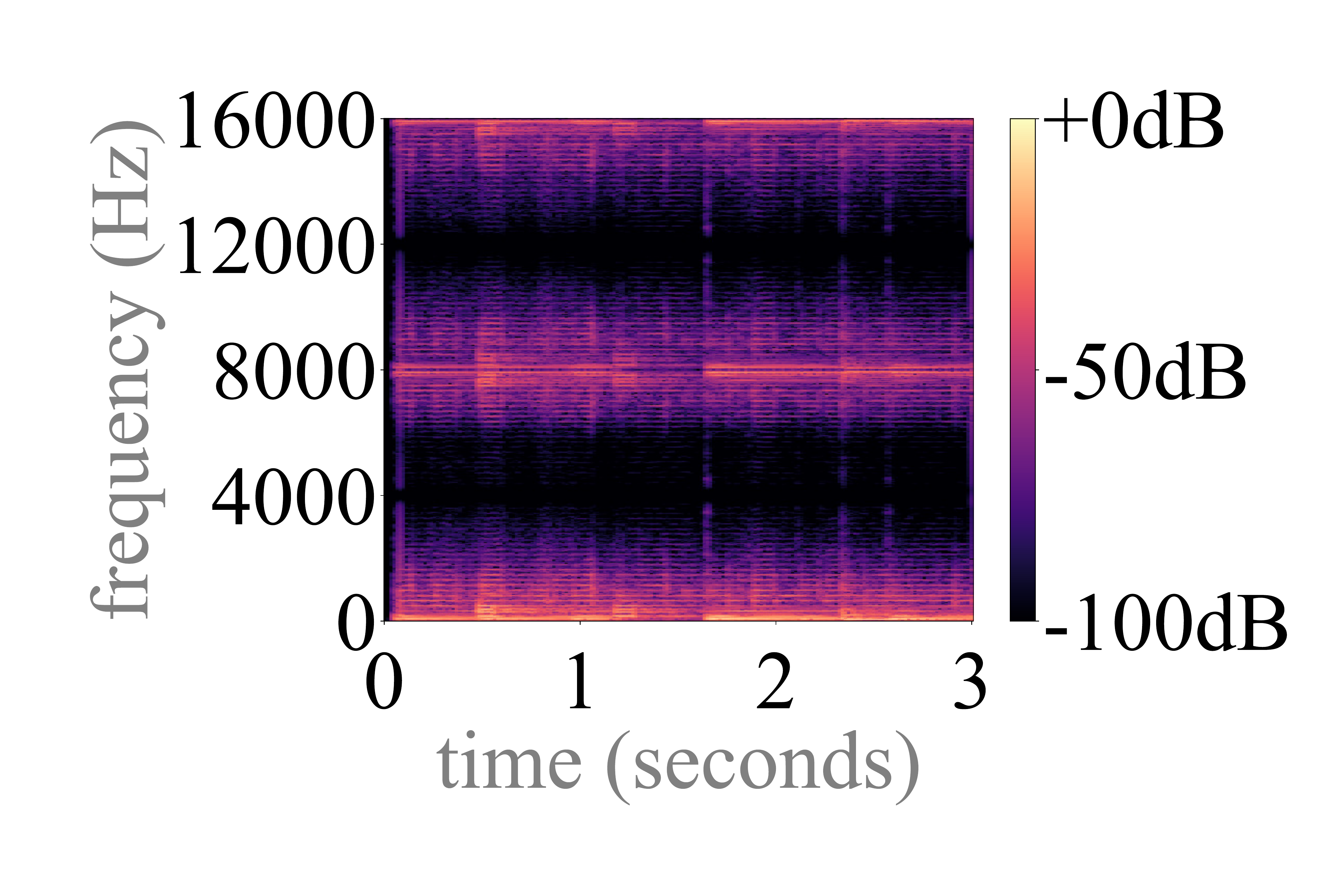}}		\vspace{-2mm}
		{(d) Subpixel CNN\\(filter length=9)}\medskip
	\end{minipage}
	\vfill \vspace{-4mm}
	\hspace{18mm}
	\begin{minipage}[b]{0.18\linewidth}
		\centering
		\centerline{\includegraphics[width=1.1\linewidth]{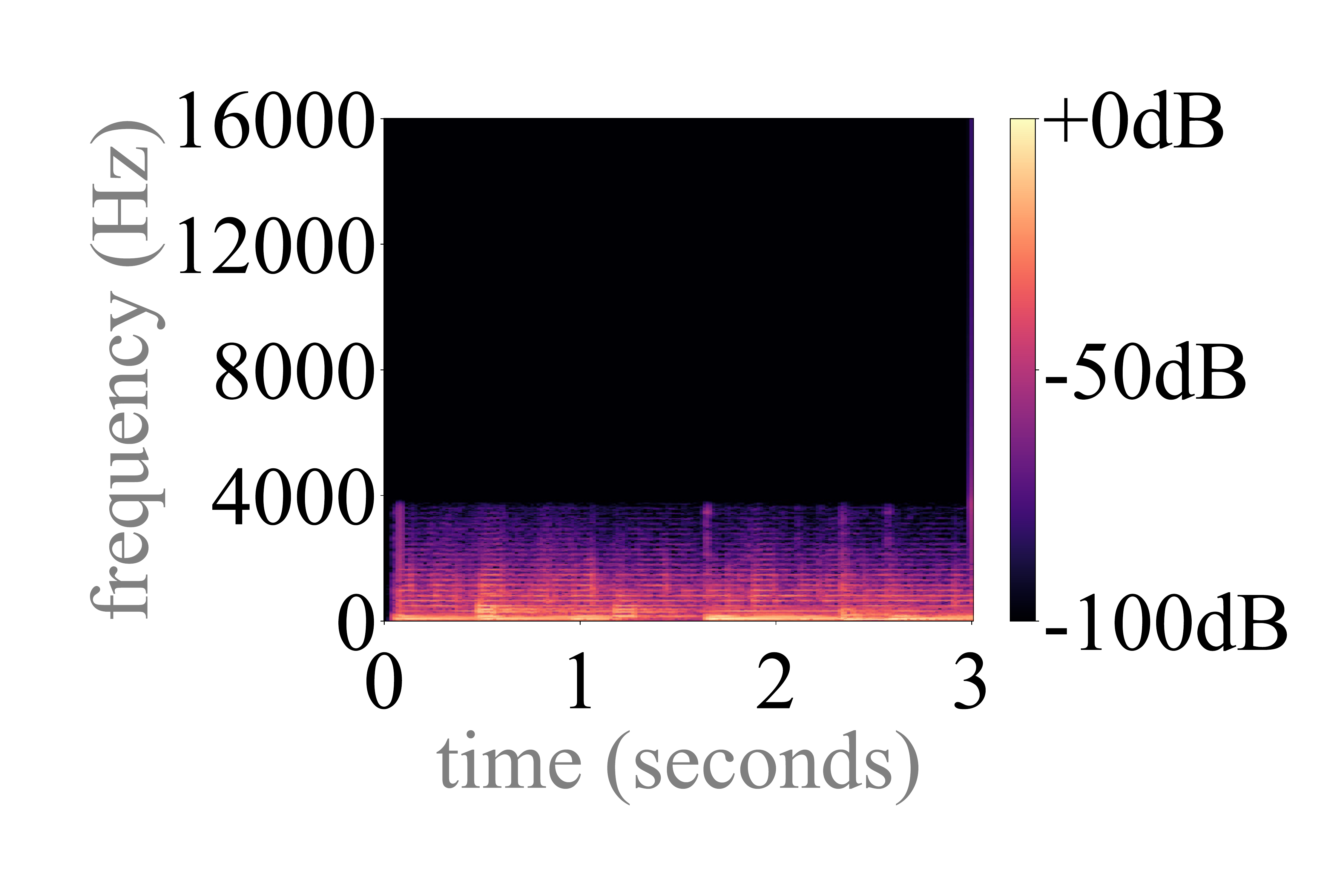}}		\vspace{-2mm}
		{(e) Interpolation:\\sinc}\medskip
	\end{minipage}	
	\begin{minipage}[b]{0.18\linewidth}
		\centering
		\centerline{\includegraphics[width=1.1\linewidth]{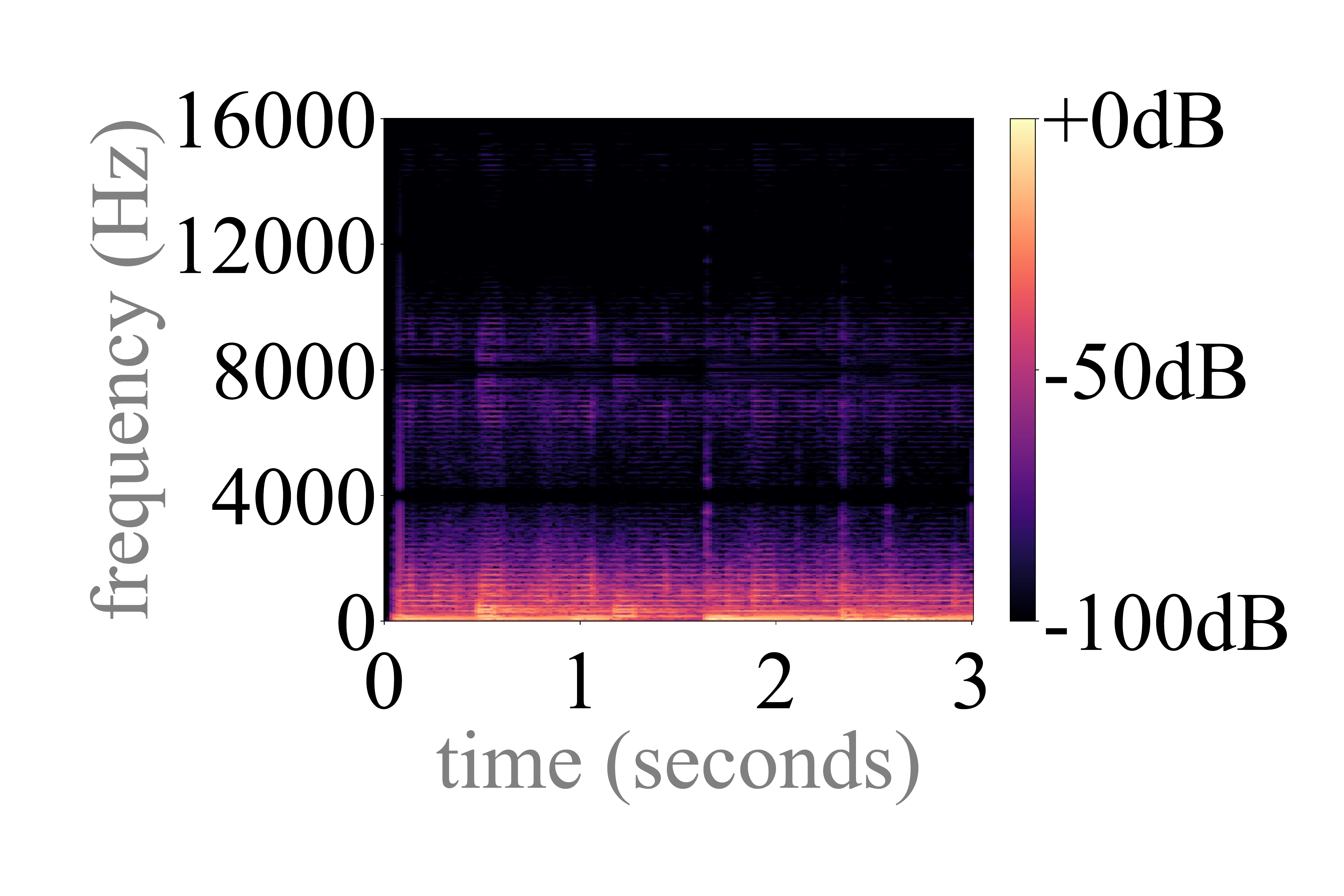}}		\vspace{-2mm}
		{(f) Interpolation:\\linear}\medskip
	\end{minipage}
	\begin{minipage}[b]{0.18\linewidth}
		\centering
		\centerline{\includegraphics[width=1.1\linewidth]{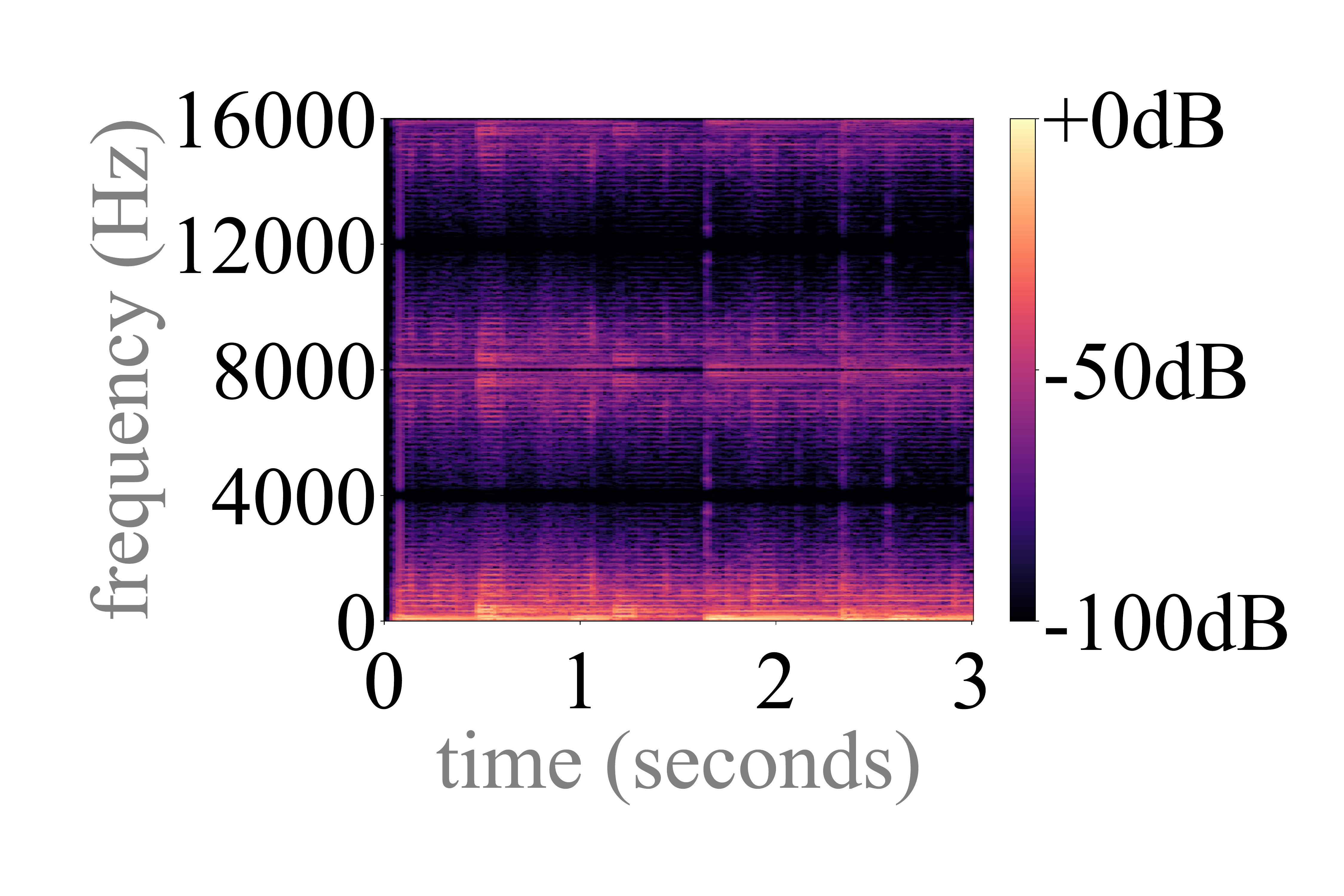}}		\vspace{-2mm}
		{(g) Interpolation:\\nearest neighbor}\medskip
	\end{minipage}
	\begin{minipage}[b]{0.18\linewidth}
		\centering
		\centerline{\includegraphics[width=1.1\linewidth]{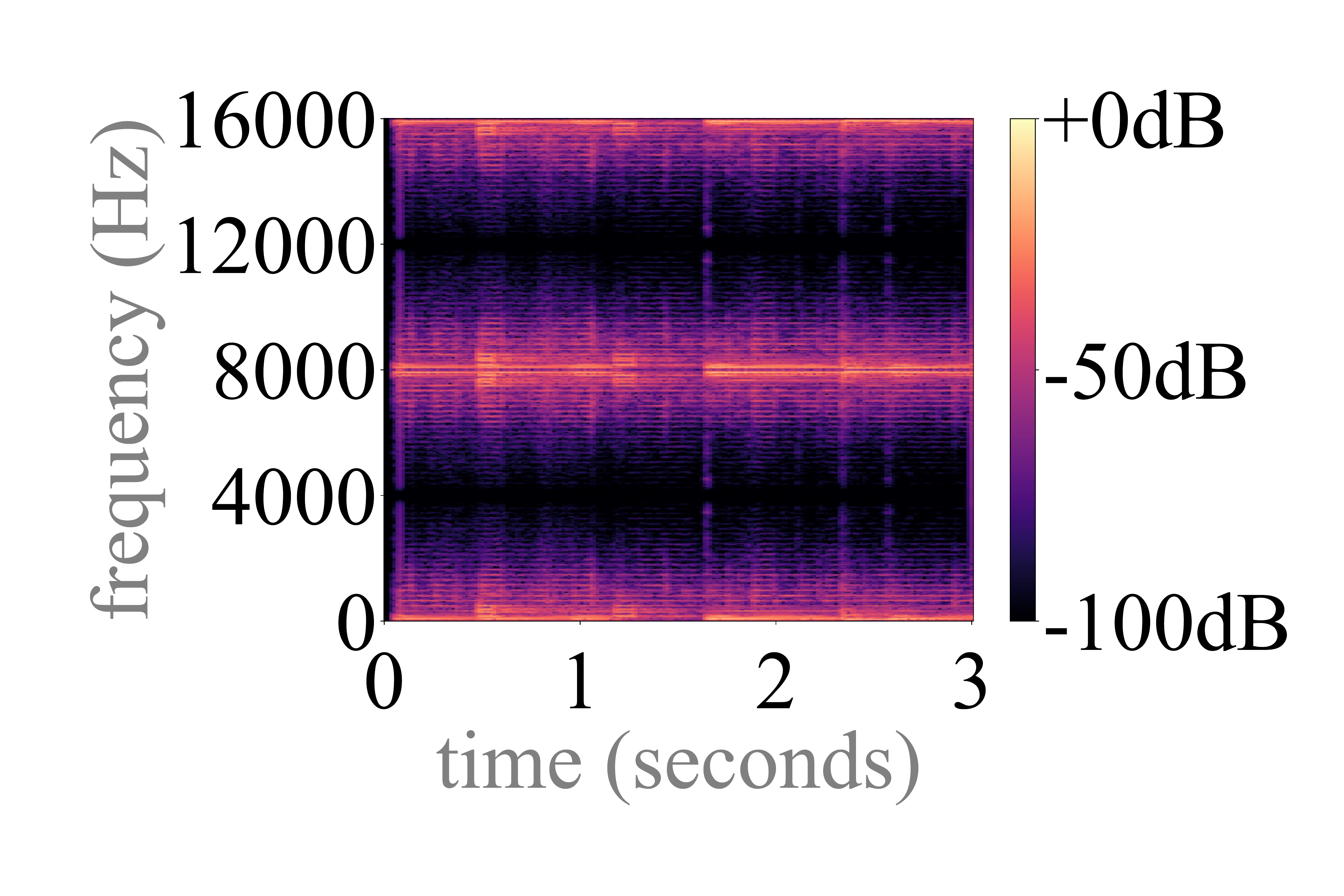}}		\vspace{-2mm}
		{(h) Interpolation:\\stretch}\medskip
	\end{minipage}
	\vfill \vspace{-5mm}
	\caption{Input: music signal at 8kHz. Upsampling~($\uparrow$4) layers can introduce artifacts and spectral replicas~(clearly depicted here). Transposed CNNs with stride=4. This figure was not included in the original manuscript due to space constrains.}
	\vspace{-6mm}
	\label{fig:all}
\end{figure*}

\end{document}